\pdfoutput=1
\documentclass[aps,twocolumn,prd,superscriptaddress,showpacs,preprintnumbers,amsmath,amssymb,nofootinbib]{revtex4-1}

\usepackage{dsfont}

\usepackage{graphicx}
\usepackage{slashed}% for slashed ...
\usepackage{hyperref}

\usepackage[T1]{fontenc}
\usepackage[latin9]{inputenc}
\usepackage{amsmath,amssymb,amsthm,amsfonts}
\usepackage{mathrsfs}
\usepackage{color,rotating}
\usepackage{dsfont}
\usepackage{setspace}
\usepackage{verbatim}
\usepackage{fancyhdr}
\usepackage{float}

\def\s0#1#2{\mbox{\small{$ \frac{#1}{#2} $}}}
\def\0#1#2{\frac{#1}{#2}}

\usepackage[caption=false]{subfig}
\graphicspath{{./figures/}}
\newcommand{\bea}{\begin{eqnarray}}
\newcommand{\eea}{\end{eqnarray}}
\definecolor{darkgreen}{rgb}{0,0.6,0}
\definecolor{gray}{rgb}{.7,.7,.7}

\newcommand {\apgt} {\ {\raise-.5ex\hbox{$\buildrel>\over\sim$}}\ }
\newcommand {\aplt} {\ {\raise-.5ex\hbox{$\buildrel<\over\sim$}}\ }

\def\s0#1#2{\mbox{\small{$ \frac{#1}{#2} $}}}
\def\0#1#2{\frac{#1}{#2}}

\newcommand{\be}{\begin{eqnarray}}
\newcommand{\ee}{\end{eqnarray}}

\newcommand{\imag}{\text{i}}
\newcommand{\skipthis}[1]{}
\renewcommand{\d}{{\text{d}}}

\begin{document}

\title{Chiral Mirror-Baryon-Meson Model and Nuclear Matter beyond Mean-Field}

  \author{Johannes Weyrich}
 \email{weyrich@theorie.ikp.physik.tu-darmstadt.de}
 \affiliation{Theoriezentrum, Institut f\"{u}r Kernphysik, Technische Universit\"{a}t Darmstadt, Schlossgartenstra\ss e 2, 64289 Darmstadt, Germany}

\author{Nils Strodthoff}
\email{n.strodthoff@thphys.uni-heidelberg.de} 
\affiliation{Institut f\"ur Theoretische
  Physik, Universit\"at Heidelberg, Philosophenweg 16, 69120
  Heidelberg, Germany} 

  \author{Lorenz von Smekal}
 \email{lorenz.smekal@physik.tu-darmstadt.de}
 \affiliation{Theoriezentrum, Institut f\"{u}r Kernphysik, Technische Universit\"{a}t Darmstadt, Schlossgartenstra\ss e 2, 64289 Darmstadt, Germany}

		%Lattice QCD calculations (see also 11.15.Ha
                        %Lattice gauge theory)
\begin{abstract}
  We consider a chiral baryon-meson model for nucleons and their parity partners in mirror assignment interacting with pions, sigma and omega mesons to describe the liquid-gas transition of nuclear matter together with chiral symmetry restoration in the high density phase. Within the mean-field approximation the model is known to provide a phenomenologically successful description of the nuclear-matter transition. Here, we go beyond this approximation and include mesonic fluctuations by means of the functional renormalization group. While these fluctuations do not lead to major qualitative changes in the phase diagram of the model, beyond mean-field, one is no-longer free to adjust the parameters so as to reproduce the binding energy per nucleon, the nuclear saturation density, and the nucleon sigma term all at the same time. However, the prediction of a clear first-order chiral transition at low temperatures inside the high baryon-density phase appears to be robust.
\end{abstract}
\maketitle
%%%%%%%%%%%%%%%%%%%%%%%%%%%%%%%%%%%%%%%%%%%%%%%%%%%%%%%%%%%%%%%%%%%%%%%%%%%%%%%%%%%%%%%%%%%%%%%%%%%%%%%%%
\section{Introduction}
Relativistic heavy-ion collisions probe the state of strong-interaction matter at finite temperature
and finite baryon density \cite{BraunMunzinger:2008tz,Fukushima:2013rx}. A particularly interesting region in the phase diagram is that of large baryon densities.
In this work we concentrate on cold nuclear matter in the vicinity of the liquid-gas transition. Unfortunately the sign problem in Lattice QCD prevents a straightforward investigation of this region of the phase diagram through Monte-Carlo simulations. Methods to circumvent this problem are being actively developed. For example, complex Langevin dynamics has recently been successfully used for full QCD at finite density albeit on still rather small lattices \cite{Sexty:2013ica}. Strong-coupling techniques can either be used together with hopping parameter expansions for very heavy quarks \cite{Fromm:2012eb,Langelage:2014vpa} or to derive a graph representation valid also for light quarks but without continuum limit \cite{deForcrand:2009dh,deForcrand:2014tha}. More direct evidence of a first-order transition analogous to that of nuclear matter has so far only been seen in $G_2$-QCD at finite density, a QCD-like theory with fermionic baryons but without sign problem which can therefore be simulated with standard lattice techniques \cite{Maas:2012wr,Wellegehausen:2013cya}.   

Meanwhile, effective models such as the Walecka model \cite{Walecka1974491,Serot:1984ey,Serot:1997xg} have been used since the 1970s and are still being used to describe the properties of nuclear matter in the low temperature region of the QCD phase diagram around the liquid-gas transition. In its simplest form the Walecka model consists of a single nucleon field coupled to a neutral scalar field and a neutral vector meson field. In the mean-field approximation this model shows nuclear-matter saturation, when adjusting the model parameters to realize the nuclear-matter binding energy $E_b\simeq16$\,MeV per nucleon together with a saturation density of $n_0=0.16$\;fm$^{-3}$. On the other hand neutron matter remains unbound in the Walecka model. %problems of the Walecka model?
In the chiral Walecka model the fundamental degrees of freedom are nucleons coupled to the scalar-pseudoscalar sector in a chirally invariant way
as well as to a neutral vector meson. Early on, however, it was noticed that on the mean-field level the chiral Walecka model leads to massless Lee-Wick matter, with massless baryonic degrees of freedom in the chirally restored phase. Nevertheless, using a phenomenological parametrization of the effective action at zero-temperature and density as the input, it has recently received renewed interest. This is because it then proved to be very useful for the investigation of nuclear and neutron matter at small temperatures both in the mean-field approximation \cite{Floerchinger:2012xd} and beyond \cite{{Drews:2013hha,Drews:2014wba,Drews:2014spa}}.

The difficulties with the chiral Walecka model at zero temperature are related to the way the nucleon mass is generated in this model, predominantly by dynamical chiral symmetry breaking as in quark-meson-models. Unlike constituent quarks, however, the nucleons are not expected to become massless when chiral symmetry gets restored. This is in fact supported by lattice studies in which the effects of chiral symmetry restoration have been studied by removing low-lying Dirac modes in valence quark propagators \cite{Glozman:2012iu,Glozman:2012fj,Glozman:2012hw}. An alternative model to reflect the fact that the nucleon mass stays
finite when chiral symmetry gets restored as expected with increasing
baryon chemical potential beyond the nuclear-matter transition must therefore include the nucleons' parity partners \cite{Cohen:2002st,Jaffe:2006jy}, conventionally identified either with the $N(1535)$ or the heavier $N(1650)$.\footnote{Here we use the $N(1535)$ but for the purpose of our present study one might as well use the $N(1640)$ with very minor changes. For a recent discussion of both assignments see, e.g., \cite{Gallas:2013ipa}.} Such models are called parity-doublet or mirror models \cite{Detar:1988kn,Hatsuda:1988mv,Jido:1998av}.

Parity-doublet models have provided very promising results already at tree-level and in mean-field (MF) approaches \cite{Jido:2001nt,Wilms:2007uc,Zschiesche:2006zj,Dexheimer:2007tn,Gallas:2009qp,Sasaki:2010bp,Sasaki:2011ff,Gallas:2011qp,Steinheimer:2011ea,Paeng:2011hy,Giacosa:2011qd,1502.05969v1}. This was the primary motivation to examine the effects of including mesonic fluctuations in such a model which is done here using the functional renormalization group (FRG). The paper is organized as follows: In Sec.~\ref{sec:theory} we review the mirror assignment, the parity-doublet model and the role of bosonic and fermionic fluctuations in this model. In Sec.~\ref{sec:results} we describe our results in the different approximation schemes at first vanishing temperature, and finally also at finite temperature before we summarize and conclude in Sec.~\ref{sec:conclusion}.

%%%%%%%%%%%%%%%%%%%%%%%%%%%%%%%%%%%%%%%%%%%%%%%%%%%%%%%%%%%%%%%%%%%%%%%%%%%%%%%%%%%%%%%%%%%%%%%%%%%%%%%%%%%%%%%%%%%%%%%%%%%%%%%%%%%%%%%%%%%%%%%%%%%%%%%%%%%%%
\section{Parity-doublet model and the nuclear matter transition}
\label{sec:theory}

%%%%%%%%%%%%%%%%%%%%%%%%%%%%%%%%%%%%%%%%%%%%%%%%%%%%%%%%%%%%%%%%%%%%%%%%%%%%%%%%
%%%%%%%%%%%%%%%%%%%%%%%%%%%%%%%%%%%%%%%%%%%%%%%%%%%%%%%%%%%%%%%%%%%%%%%%%%%%%
\subsection{Mirror assignment and chirally invariant mass}

%mirror assignment
First recall the possible representations for baryons under chiral $SU(2)_L\times SU(2)_R$ transformations by considering the product of three $(\frac{1}{2},0)\oplus(0,\frac{1}{2})$ quark representations where one assigns, without loss, say the first label in the product representations $(m,n)$ to an irreducible representation of $SU(2)_L$ and the second to one of $SU(2)_R$,
yielding \cite{Jido:2001nt},
\begin{align}
&\left((\tfrac{1}{2},0)\oplus(0,\tfrac{1}{2})\right)\otimes\left((\tfrac{1}{2},0)\oplus(0,\tfrac{1}{2})\right)\otimes\left((\tfrac{1}{2},0)\oplus(0,\tfrac{1}{2}))\right)\nonumber \\[4pt]
  &\hspace{1cm }=5 \left(\tfrac{1}{2},0)\oplus(0,\tfrac{1}{2})\right)
  \oplus  3 \left((1,\tfrac{1}{2})\oplus(\tfrac{1}{2},1)\right)\\[4pt]
 &\hspace{5cm}  \oplus \left((\tfrac{3}{2},0)\oplus (0,\tfrac{3}{2})\right)\, .
 \nonumber
\end{align}
For the iso-doublet of proton and neutron one thus of course uses the same
$(\frac{1}{2},0)\oplus(0,\frac{1}{2})$ representation as for the quarks, for left and right-handed Dirac fermions $\psi_{l/r}\equiv \frac{1\mp\gamma_5}{2}\psi$,  so that the nucleon fields transform just like the pair of up and down quark fields. 

If one now considers two baryon species $\psi_1$ and $\psi_2$, both in a  $(\frac{1}{2},0)\oplus(0,\frac{1}{2})$ representation, there are two possibilities:
Assigning the first label of the first species say to $\psi_{1,l}$ as before, the first label of the second species can either refer to $\psi_{2,l}$ or to $\psi_{2,r}$. The first assignment, where both fermions transform exactly in the same way, is called the {\em naive} assignment whereas the second assignment, where the transformation properties of the second left/right-handed fermion species under chiral transformations are interchanged relative to the first, is referred as the {\em mirror} assignment. These two assignments are the only possibilities in a two-baryon system, provided both species belong to a $(\frac{1}{2},0)\oplus(0,\frac{1}{2})$ representation. Explicitly, they correspond to the transformation properties
\begin{equation}
\begin{split}
\psi_{1,l} &\rightarrow \Omega_{L}\psi_{1,l},\hspace{1mm}\psi_{1,r} \rightarrow\Omega_{R}\psi_{1,r}\\
\psi_{2,l} &\rightarrow \Omega_{L}\psi_{2,l},\hspace{1mm}\psi_{2,r} \rightarrow\Omega_{R}\psi_{2,r}
\end{split}
\end{equation} 
for the naive assignment, and
\begin{equation}
\begin{split}
\psi_{1,l} &\rightarrow \Omega_{L}\psi_{1,l},\hspace{1mm}\psi_{1,r} \rightarrow\Omega_{R}\psi_{1,r}\\
\psi_{2,r} &\rightarrow \Omega_{L}\psi_{2,r},\hspace{1mm}\psi_{2,l} \rightarrow\Omega_{R}\psi_{2,l}.
\end{split}
\end{equation} 
for the mirror assignment, where $\Omega_{R,L}\in SU(2)_{R,L}$.

%chiral invariant interaction terms
Obviously, the generic kinetic term
\begin{equation}\label{kineticterm}
\mathcal L_\text{kin} = \sum_{i={1,2}} \bar{\psi}_i\imag\slashed{\partial}\psi_i
\end{equation}
is invariant under chiral transformations, irrespective of the assignment.
Conversely, for both assignments a conventional Dirac-mass term
\begin{equation}\label{massterm}
\mathcal L_m = -m\bar{\psi}_i\psi_i
\end{equation}
explicitly breaks chiral symmetry in either case.
Hence, it is impossible to write down a chirally invariant
mass term for a single Dirac-fermion species. Interestingly, however,
this is no longer true for two fermion species in the mirror assignment. In this case a mass term of the form 
\begin{equation}
\label{mirrormassterm}
\begin{split}
\mathcal L_{m,\text{mirror}}=&m_0(\bar{\psi}_2\psi_1+\bar{\psi}_1\psi_2)\\
=& m_0(\psi^{\dag}_{2r}\psi_{1l}+\psi^{\dag}_{1l}\psi_{2r}+\psi^{\dag}_{1r}\psi_{2l}+\psi^{\dag}_{2l}\psi_{1r})
\end{split}
\end{equation}
remains invariant under the full chiral  $SU(N_f)_L\times SU(N_f)_R$.
In the naive assignment such a term would of course also break chiral symmetry. The particularly appealing feature of the mirrror model thus is the possibility of having a  chirally invariant local mass term.

%%%%%%%%%%%%%%%%%%%%%%%%%%%%%%%%%%%%%%%%%%%%%%%%%%%%%%%%%%%%%%%%%%%%%%%%%%%%%%%%%%%%%%%%%%%%%%%%%%%%%%%%%%%%%%%%%%%%%%%%%%%%%%%%%%%%%%%%%%%%%%%%%%%%%%%%%%%%%
\subsection{Construction of the parity-doublet model}
The parity-doublet (or mirror) model consists of two species of mirror-assigned baryons with a chirally invariant mass term of the form of Eq.~(\ref{mirrormassterm}). Compared to the previous subsection they are introduced as $N_{1} \equiv \psi_1$ and $N_2\equiv \gamma_5 \psi_2$ so that $N_2$ has the opposite parity of $N_1$ and the eigenvalues of the mass matrix in the chiral limit are both $m_0$, rather than $\pm m_0$ for that in Eq.~(\ref{mirrormassterm}), see  \cite{Jido:2001nt}. Both baryon species are coupled to the scalar/pseudo-scalar  meson sector $\vec\phi =(\sigma,\vec\pi)$ in an $SO(4)$-invariant way.
The corresponding Euclidean Lagrangian (including baryon chemical potential $\mu_B$ and a vector coupling to the $\omega$-meson) then reads
\begin{equation}
\label{eq:Lagrangian}
\begin{split}
\mathcal{L}=&\bar N_1  \big(\slashed \partial-\mu_B \gamma_0 + h_1(\sigma +\imag \gamma_5 \vec \tau \vec \pi)+ \imag h_v \gamma^\mu\omega_\mu \big) N_1\\
&+\bar N_2  \big(\slashed \partial-\mu_B \gamma_0 + h_2(\sigma -\imag \gamma_5 \vec \tau \vec \pi)+ \imag h_v \gamma^\mu\omega_\mu \big) N_2\\
&+m_0 (\bar N_1 \gamma_5 N_2 - \bar N_2 \gamma_5 N_1)+\mathcal L_\text{mes}.\\
\end{split}
\end{equation}
Note that the mirror baryon $N_2$ transforms as $N_2\to e^{-i\theta^a \gamma^5 T^a}N_2$ under axial transformations whereas the original baryon
transforms as $N_1\to e^{i\theta^a \gamma^5 T^a}N_1$, which requires opposite signs in their Yukawa couplings to the pion. Furthermore, we have assumed
the same strength $\imag h_v$  for the  Yukawa coupling of both to the iso-scalar vector meson which is imaginary here, in order to describe a repulsive four-fermion vector interaction. The mesonic part is given by 
\begin{equation}
\mathcal{L}_\text{mes}=\frac{1}{2}\partial_\mu\vec \phi\partial^\mu \vec \phi+\frac{1}{4}F_{\mu\nu}F^{\mu\nu}+U(\phi^2,\omega^2)-c\sigma\,,
\end{equation}
where $F^{\mu\nu}=\partial^\mu \omega^\nu- \partial^\nu \omega^\mu$ and the mesonic potential at tree-level, i.e. in the microscopic bare action at the ultraviolet cutoff scale $\Lambda$, or in the mean-field approximation is of the form
\begin{equation}
 U(\phi^2,\omega^2) = -\frac{\mu^2}{2}\phi^2+\frac{\lambda}{4}\phi^4+\lambda_6\phi^6 + \frac{m_\omega^2}{2} \omega^2 , \label{barepot}
\end{equation}
with $\phi^2=\sigma^2+\vec{\pi}^2$ and parameters $\mu^2,\lambda,\lambda_6$. A non-vanishing pion mass  is taken into account by means of an explicit linear breaking term $c \sigma$, i.e.\ 
\begin{equation}
	 V_{\text{M}} = U(\phi^2,\omega^2 )-c\sigma.
\end{equation}

The vector meson could in principle be consistently included as a fluctuating field, for example in the Stueckelberg formalism \cite{Urban:2001ru,Struber:2007bm}, 
but with $m_\omega = 782$ MeV its mass is comparatively large so that it will not have much of an impact as a fluctuating meson field in loops. Instead, we therefore simply evaluate its expectation value on a given solution for the effective potential in a mean-field treatment in the following. Due to the repulsive nature of the vector interaction, this expectation value will be purely imaginary  corresponding to a stationary phase approximation for complex saddle points.  

Rewriting the Lagrangian in momentum space we find
\begin{equation}
\mathcal{L}=\bar \Psi S_0^{-1} \Psi +\mathcal{L}_\text{mes},
\end{equation}
where $\Psi=\left(\begin{smallmatrix}N_1\\N_2\end{smallmatrix}\right)$ and
\begin{align}
\label{eq:s0m1}
&S_0^{-1}=\\
&\left(\begin{smallmatrix}-\imag \slashed p +h_1(\sigma+\imag \gamma_5 \vec \pi \vec \tau)+ \imag h_v \slashed\omega-\mu_B\gamma^0& m_0 \gamma_5\\ -m_0 \gamma_5&-\imag \slashed p +h_2(\sigma-\imag \gamma_5 \vec \pi\vec \tau)+\imag h_v \slashed \omega-\mu_B\gamma^0\end{smallmatrix}\right).\nonumber
\end{align}
%%%%%%%%%%%%%%%%%%%%%%%%%%%%%%%%%%%%%%%%%%%%%%%%%%%%%%%%%%%%%%%%%%%%%%%%%%%%%%%%%%%%%%%%%%%%%%%%%%%%%%%%%%%%%%%%%%%%%%%%%%%%%%%%%%%%%%%%%%%%%%%%%%%%%%%%%%%%%%
\subsection{Fermionic and bosonic fluctuations in the parity-doublet model}
%%%%%%%%%%%%%%%%%%%%%%%%%%%%%%%%%%%%%%%%%%%%%%%%%%%%%%%%%%%%%%%%%%%%%%%%%%%%%%%%%%%%%%%%%%%%%%%%%%%%%%%%%%%%%%%%%%%%%%%%%%%%%%%%%%%%%%%%%%%%%%%%%%%%%%%%%%%%%%
%\subsubsection{Mean-field Analysis}
In the mean-field approximation the inverse baryon propagator is given by
\begin{equation}
S_{0}^{-1}=\left( \begin{smallmatrix}-\imag \slashed p +h_1 \sigma-\tilde\mu_B& m_0 \gamma_5\\ -m_0 \gamma_5&-\imag \slashed p +h_2 \sigma-\tilde \mu_B\end{smallmatrix}\right),
\end{equation}
where $\tilde \mu_B=\mu_B- \imag h_v \omega_0$ is the baryon chemical potential shifted by a potentially non-zero value of the zero-component of the $\omega$-meson field, with $\Delta\mu_B = \imag h_v\omega_0 \ge 0$, in the rest-frame of an isotropic thermal medium in which the expectation values of the spatial components vanish. The determinant required for the computation of the grand potential is given by
\begin{equation}
\begin{split}
%\det \gamma_0 S_{0}^{-1}&=\left[m_0^4+2 m_0^2(-(\mu+\imag p_0)^2+\vec p^2+h_1h_2 \sigma^2)\right.\\
%$&\quad\left.+((\tilde \mu_B+\imag p_0)^2-\vec p^2-h_1^2\sigma^2)((\tilde \mu_B+\imag p_0)^2-\vec p^2-h_2^2\sigma^2)\right]^2\\
\det \gamma_0 S_{0}^{-1}\equiv \left((p_0-\imag \tilde \mu_B)^4+\alpha_{p}\, (p_0-\imag \tilde \mu_B)^2 +\beta_{p} \right)^2\,
\end{split}
\end{equation}
where
\begin{equation}
\begin{split}
\alpha_{p} &=2 m_0^2 +2 \vec p^{\,2}+h_1^2\sigma^2+h_2^2 \sigma^2,\\
\beta_{p} &= \frac{\alpha^2_{p}}{4} - \frac{1}{4} (h_1-h_2)^2\sigma^2(4 m_0^2+(h_1+h_2)^2\sigma^2)\,.
\end{split}
\end{equation}
Its zeros $m_\pm=\imag p_0$ (at vanishing 3-momentum $\vec p=0$ and chemical potential $\tilde \mu_B=0$) define the mean-field baryon masses, which are given by
\begin{equation}
m_\pm=\frac{1}{2}\left(\pm(h_1-h_2)\sigma +\sqrt{4 m_0^2+(h_1+h_2)^2\sigma^2}\right) \label{eq:mirrormasses}
\end{equation} 
In particular, in the chirally symmetric case for $\sigma=0$ the baryons become degenerate with mass $m_0$ again. Evaluating the mean-field grand potential for a sharp momentum cutoff leads to
\begin{equation}
  \Omega= \Omega_q + V_\text{M} = -T \sum_{p_0}\int\frac{\d^3 p}{(2\pi)^3} \text{Tr} \log S_0^{-1}+V_{\text{M}}
\end{equation}
with
\begin{align}
\Omega_q &=-2N_f \sum_\pm\int\frac{\d^3 p}{(2\pi)^3} \log\left(\cosh\left(\frac{\epsilon_p^\pm+\tilde \mu_B}{2T}\right)\right) \nonumber \\ 
&=-2 N_f \sum_\pm\int\frac{\d^3 p}{(2\pi)^3}\bigg[\frac{|\epsilon_p^\pm+\tilde \mu_B|}{2}   +\frac{|\epsilon_p^\pm-\tilde \mu_B|}{2}\\
  &+T \log\bigg(1+e^{-\frac{|\epsilon_p^\pm+\tilde \mu_B|}{T}}\bigg)
  +T \log\bigg(1+e^{-\frac{|\epsilon_p^\pm-\tilde \mu_B|}{T}}\bigg)\bigg]\, , \nonumber
\end{align}
where we have introduced single-quasiparticle energies 
\begin{equation}
\label{eq:defepsilon}
\epsilon_p^\pm=\sqrt{\frac{\alpha_p}{2}\pm \sqrt{\frac{\alpha^2_p}{4}-\beta_p}}\,.
\end{equation}
For $m_0=0$ one simply has $\epsilon_p^\pm= \sqrt{ \vec p^{\,2}+ m_\pm^2 }$.
The expression for the fermionic part of the grand potential then reduces to a sum of two usual ones, as e.g., in quark-meson models, with masses $m_+ = h_1\sigma$ and $m_- = h_2\sigma$, respectively. The number of flavors here is $N_f=2$ representing two iso-doublets, one for the nucleons $(p,n)$ and one for their parity partners.

The mean-field calculation can be improved by including mesonic fluctuations, which is done here within the framework of the Functional
Renormalization Group (FRG), a powerful non-perturbative tool in quantum field theory and statistical physics, see \cite{Berges:2000ew,Pawlowski:2005xe,Gies:2006wv,Schaefer:2006sr,Braun:2011pp} for QCD-related reviews. The central object in the approach pioneered by Wetterich is the effective average action $\Gamma_k$ which generalizes the effective action $\Gamma$ by introducing a coarse graining scale up to which quantum fluctuations are included. On a technical level this is achieved by means of a regulator function $R_k$ which acts like a mass term in the IR. The RG-scale $k$ is taken from the ultraviolet (UV) cutoff scale $\Lambda$ down to zero and in turn the effective average action interpolates between a microscopic bare action at $\Lambda$ and the effective action $\Gamma$ at $k=0$. The evolution of the effective average action with $k$ is described by an exact 1-loop equation, however involving full field- and scale-dependent propagators, which takes the form
\begin{equation}\label{eq:fullwettericheq}
\partial_k \Gamma_{k}= \tfrac{1}{2}\operatorname{Tr}\bigg[\frac{1}{\Gamma_{k}^{(2)}+R_{k}}\partial_k R_{k}\bigg],
\end{equation}
where $\Gamma_k^{(2)}$ is the second functional derivative of the effective average action with respect to the fields. The trace includes a momentum integration as well as traces over all inner indices.

To evaluate (\ref{eq:fullwettericheq}) one has to specify an Ansatz for the effective action. Here we work in the local potential approximation (LPA) where the only scale-dependence stems from the effective potential, and correspondingly we use
\begin{equation}
\Gamma_k=  \int d^4x \;  \mathcal{L}|_{U\to U_k}
\end{equation}
with the Lagrangian from Eq.~(\ref{eq:Lagrangian}) starting with a bare potential $U_\Lambda$ at $k=\Lambda$ of the form as given in Eq.~(\ref{barepot}). The flow for the effective potential decomposes into fermionic 
and bosonic contributions,
\begin{equation}
\partial_k U_k=\partial_k U_{k,F}+\partial_k U_{k,B}\,.
\end{equation} 
Using the $3d$-analogue of the LPA-optimized regulator from Ref.~\cite{Litim:2001up},
\begin{equation}
	R_{k,F}(\vec{p}) = -\mathrm i\vec{p}\cdot\vec{\gamma}\left(\sqrt{\tfrac{k^2}{\vec{p}^2}}-1\right)
	\theta(k^2-\vec{p}^2)\,,
\end{equation}  
the fermionic contributions to the flow of the effective potential can be obtained analogous to that of the quark-meson-diquark model for two-color QCD in \cite{Strodthoff:2011tz}, giving 
\begin{equation}\label{eq:fermionmirrorflow}
\begin{split}
\partial_k U_{k,F}&=-\frac{N_f k^4}{6\pi^2}\sum_\pm\bigg[\frac{2(k^2+m_0^2-\epsilon_k^\pm {}^2)+(h_1^2+h_2^2)\sigma^2}{(\epsilon_k^\mp {}^2 -\epsilon_k^\pm{}^2)\epsilon_k^\pm}\\
&\times\left(\tanh\left(\tfrac{\epsilon_k^\pm+\tilde\mu_B}{2T}\right)+\tanh\left(\tfrac{\epsilon_k^\pm-\tilde\mu_B}{2T}\right)\right)\bigg]\,,
\end{split}
\end{equation}
where we have again used the single-quasiparticle energies $\epsilon_k^\pm$ defined in (\ref{eq:defepsilon}), here at $\vec p^{\, 2} = k^2$. As in the mean-field approximation, for
$m_0=0$ the two fermion species decouple and (\ref{eq:fermionmirrorflow}) reduces to the sum of two quark-meson model contributions to the flow of the effective potential.

The bosonic contribution to the flow of the effective potential is identical
to the expression in quark-meson models \cite{Schaefer:2006sr} and reads, for the $3d$-analogue of the LPA-optimized regulator,     
\begin{equation}  
       \partial_{k}U_{k,B}=
      \frac{k^4}{12\pi^2}\bigg[\frac{1}{\epsilon_k^{\sigma}}\coth\left(\frac{\epsilon_k^{\sigma}}{2T}\right)
      +
      \frac{3}{\epsilon_k^{\pi}}\coth\left(\frac{\epsilon_k^{\pi}}{2T}\right)\bigg], \label{eq:bosflow}
\end{equation}
where one introduces mesonic single-quasiparticle energies via
      \begin{equation}
      \begin{split}
\epsilon_k^{\sigma}&=\sqrt{k^2+2U_k^{\prime}+4U_k^{\prime\prime}\phi^2}\,,\\	
\epsilon_k^{\pi}&=\sqrt{k^2+2U_k^{\prime}}\,,
      \end{split}
      \end{equation}
with the notations $U_k^{\prime}=\frac{\partial}{\partial \phi^2}U_k$ and $U_k^{\prime\prime}=\frac{\partial^2}{\partial \phi^2\partial \phi^2}U_k$.
%%%%%%%%%%%%%%%%%%%%%%%%%%%%%%%%%%%%%%%%%%%%%%%%%%%%%%%%%%%%%%%%%%%%%%%%%%%%%%%%%%%%%%%%%%%%%%%%%%%%%%%%%%%%%%%

%%%%%%%%%%%%%%%%%%%%%%%%%%%%%%%%%%%%%%%%%%%%%%%%%%%%%%%%%%%%%%%%%%%%%%%%%%%%%%%%%%%%%%%%%%%%%%%%%%%%%%%%%%%%%%%%
%%%%%%%%%%%%%%%%%%%%%%%%%%%%%%%%%%%%%%%%%%%%%%%%%%%%%%%%%%%%%%%%%%%%%%%%%%%%%%%%%%%%%%%%%%%%%%%%%%%%%%%%%%%%%%%%
\subsection{Parameter fixing and phenomenology of the nuclear matter transition}
%%%%%%%%%%%%%%%%%%%%%%%%%%%%%%%%%%%%%%%%%%%%%%%%%%%%%%%%%%%%%%%%%%%%%%%%%%%%%%%%%%%%%%%%%%%%%%%%%%%%%%%%%%%%%%%%
The most relevant degrees of freedom to describe the phase diagram of strong-interaction matter near the nuclear matter transition are collective 
mesonic and baryonic excitations.
Since there are no baryons in the vacuum one should in principal fix the model
parameters at $T=0$ with a value of the baryon chemical potential close to the onset of nuclear matter where one has small nucleonic excitation energies.
Due to the Silver-Blaze property one can in principle equally well fix the model parameters at $T=\mu_B=0$, however. In the LPA this is known to introduce a slight artificial $\mu_B$-dependence, and hence a Silver-Blaze problem, mainly because the curvature masses extracted from the mesonic effective potential are not exactly the physical ones. Calculating mesonic two-point functions in random-phase approximation or from their own flow equations \cite{Strodthoff:2011tz,Kamikado:2012bt,Kamikado:2013sia,Tripolt:2013jra,Tripolt:2014wra} one observes that the physical pole masses can differ quite significantly from the curvature masses.
In the quark-meson-diquark model for two-color QCD or the quark-meson model for  QCD at finite isospin density one verifies at mean-field level that the RPA-pole masses agree with the onset of Bose-Einstein condensation of diquarks or charged pions, respectively, as they must. 
In contrast, one then deduces that especially the pion curvature mass can deviate from this by up to 30\%  \cite{Strodthoff:2011tz,Kamikado:2012bt}. Beyond mean-field, pole masses in present truncations for two-point functions are typically considerably closer to such onsets than curvature masses as well.
To completely resolve this discrepancy, and to reduce the unnaturally large differences between curvature and pole masses, one has to include wave-function renormalization beyond the LPA \cite{Helmboldt:2014iya}. 

In the parity-doublet model these differences are much smaller, however, because the baryons are much heavier than the quarks in quark-meson models. Even in the chirally symmetric regime the mirror baryons have their explicit mass $m_0$ which is of the order of 800 MeV and implies that their fluctuations are suppressed below RG scales $k \sim 800$ MeV. A manifestation of this is that we have to start the flow already in the broken phase when using a typical UV cutoff scale $\Lambda$ of about 1 GeV. Unlike the fermionic fluctuations in quark-meson models, the baryonic fluctuations of the parity-doublet model alone, between the UV cutoff and $m_0$, are thus not strong enough to drive the system into the broken phase. The mesonic fluctuations dominate the flow at $T=\mu=0$. In purely mesonic $O(N)$-models, on the other hand, the difference between pole and curvature masses is known to be a few percent effect and hence negligible \cite{Svanes:2010we,Kamikado:2013sia}. We therefore neglect this discrepancy here as well and use the standard curvature masses to fix the parameters in our calculations. In addition, we have verified, however, that a larger value for the pion curvature mass, as a simple fix to compensate missing wave-function renormalization in the vacuum, does not change our results in any substantial way.

From the effective potential in the IR one easily extracts meson curvature masses as
          \begin{equation}
            m_\sigma=\sqrt{2U_k^{\prime}+4U_k^{\prime\prime} \phi^2}\, ,\;\;
            m_{\pi}=\sqrt{2 U_k^{\prime}}\, , \;\; k\to 0 \,.
          \end{equation}
The parameters in the UV potential $V_M$ are adjusted to realize the physical pion mass $m_\pi=138$\;MeV in the IR, which essentially determines the parameter $c$.
Due to its nature as a broad resonance the mass of the sigma meson is not so well constrained, we fix it to reasonable values of $m_\sigma \simeq 500$\;MeV.

As usual, the Goldberger-Treiman relation is used to connect the minimum of the mesonic potential in the vacuum at $\vec\pi =0$, $\sigma = \bar\sigma_0$ to the pion decay constant, $\bar\sigma_0 =f_{\pi}=93$\;MeV.
The Yukawa couplings $h_{1}$ and $h_2$ are then fixed from Eq.~(\ref{eq:mirrormasses}) for a given $m_0$ by the masses $m_\pm$ of the nucleon (939\;MeV) and its parity partner (1535\;MeV).

For determining the in-medium condensate \cite{Sasaki:2010bp,1201.0950v1,1204.4318v1} at the phase transition one uses the Feynman-Hellmann theorem \cite{PhysRevC.45.1881} in combination with the Gell-Mann-Oakes-Renner relation. This provides a connection between the saturation density of nuclear matter $n_0$ and the in-medium chiral condensate $\bar \sigma(n_0)$
\begin{equation}\label{inmedium}
\frac{\bar \sigma(n_0)}{\bar \sigma_0}=1-\frac{\sigma_N}{m^2_\pi f_\pi^2}\,  n_0.
\end{equation}
Of course, the ratio $\frac{\bar \sigma(n_0)}{\bar \sigma_0}$ also depends on the value of the so-called nucleon sigma term $\sigma_N$.
The remaining model parameters $m_0$ and $h_v$ can be used to fix the right saturation density of symmetric nuclear matter $n_0\simeq0.16$\;fm$^{-3}$ and a phenomenologically reasonable value for the nucleon sigma term  $\sigma_N\simeq36$\;MeV \cite{1204.4318v1}.

%%%%%%%%%%%%%%%%%%%%%%%%%%%%%%%%%%%%%%%%%%%%%%%%%%%%%%%%%%%%%%%%%%%%%%%%%%%%%%%%%%%%%%%%%%%%%%%%%%%%%%%%%%%%%%%%
%%%%%%%%%%%%%%%%%%%%%%%%%%%%%%%%%%%%%%%%%%%%%%%%%%%%%%%%%%%%%%%%%%%%%%%%%%%%%%%%%%%%%%%%%%%%%%%%%%%%%%%%%%%%%%%%

\section{Results}
\label{sec:results}
\subsection{Extended mean-field}
We first discuss our results from the extended mean-field (eMF) approximation which  amounts to only taking the fermionic contributions to the FRG flow into account. This is comparable to the mean-field approximation provided that the contributions from the vacuum term are properly included in the grand potential \cite{Strodthoff:2011tz}.

Due to the fermionic minus sign, the fermionic contribution to the flow generally generates a negative mesonic mass term and hence drives the $\sigma$-field away from zero and into the broken phase. The integrated fermionic flow in the parity-doublet model thereby decreases more rapidly with $\sigma $ as compared to quark-meson model calculations. This is because the baryon masses start out with a rather large chirally invariant mass $m_0 $ for $\sigma =0$ already and further increase with $\sigma $, cf.~Eq.~(\ref{eq:mirrormasses}). Starting our flow at $\Lambda = 1$ GeV the baryonic fluctuations therefore
get suppressed rapidly with $\sigma $ which results in a comparatively large negative contribution to the curvature of the effective potential near the origin in field space at $\sigma =0 $. With the infrared minimum fixed at  $\bar\sigma_0 = f_\pi$ it then turns out that we have to include a small $\lambda_6 \phi^6$ term in order to generate an overall infrared potential that is sufficiently shallow around  $\bar\sigma_0$, for a sufficiently low curvature mass of the $\sigma$ meson around 500~MeV. This problem can be avoided by starting the flow at a larger UV scale $\Lambda $. We have verified that the value of $\lambda_6$ for constant infrared parameters indeed decreases with increasing $\Lambda$.  The same is true for the UV potential in the microscopic bare action when we include the mesonic fluctuations which counteract the fermionic ones. In the parity-doublet model, where the latter are not so strong, we start at $\Lambda = 1$~GeV just in the broken phase already, i.e.~with a small negative mass term in the UV potential as mentioned above.
For $\sigma$ meson masses around 500 MeV we then occasionally also need a small irrelevant coupling  $\lambda_6$ in the UV potential, if $\Lambda $ is not sufficiently large relative to $m_0$.  

\begin{figure}[H]
\centering
\includegraphics[width=0.47 \textwidth]{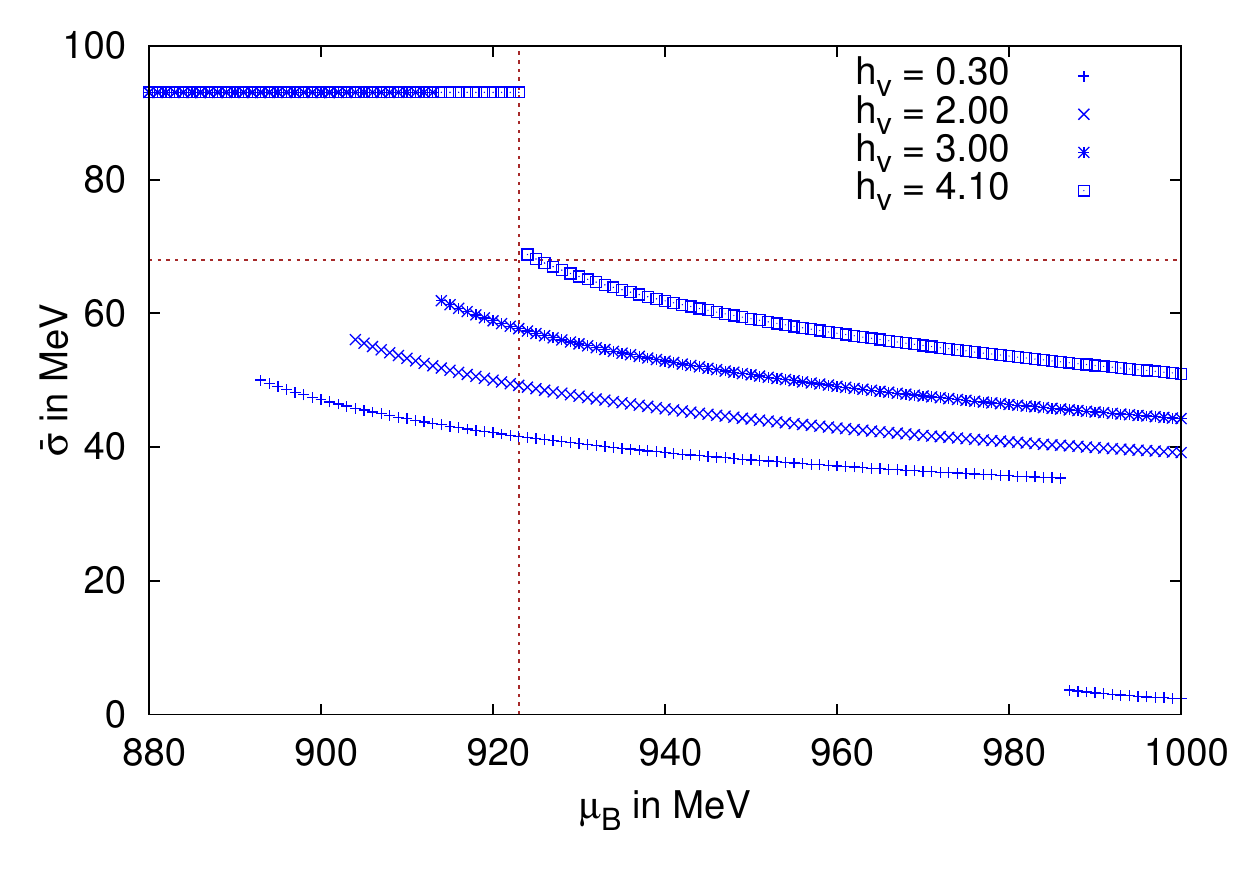}
\caption{(Color online) The chiral condensate for $m_0=820$\;MeV as a function of baryon chemical potential $\mu_B$. The plot focusses on the nuclear-matter transition and shows results for different vector couplings $h_v$. Increasing  values of $h_v$ shift the transition to larger $\mu_B$. The binding energy per nucleon (vertical dotted line), the in-medium condensate (horizontal dotted line) and the nuclear saturation density of symmetric nuclear matter are reproduced for $h_v = 4.1$.}\label{emfcondensate1}
\end{figure}

Our eMF results for the zero-temperature chiral condensate $\bar\sigma$ over the baryon chemical potential $\mu_B$ in the region of the nuclear-matter transition with  $m_0=820$\;MeV and different vector couplings $h_v$  are shown in Fig.~\ref{emfcondensate1}. The parameter set with $h_v=4.10$ 
reproduces the nuclear-matter binding energy of $E_b\simeq16$\;MeV per nucleon
(by the location of the discontinuity at $\mu_B^c=(939-16)$\;MeV$=923$\;MeV marked by the dotted red vertical line) together with a saturation density of $n_0\simeq0.16$\;fm$^{-3}$, an in-medium condensate $\bar\sigma(n_0)\simeq 69$\;MeV (marked by the dotted red horizontal line), corresponding to a nucleon-sigma term of $\sigma_N\simeq 36$\;MeV, cf.~Eq.~(\ref{inmedium}). 

\begin{figure}[h]
\centering
\includegraphics[width=0.47 \textwidth]{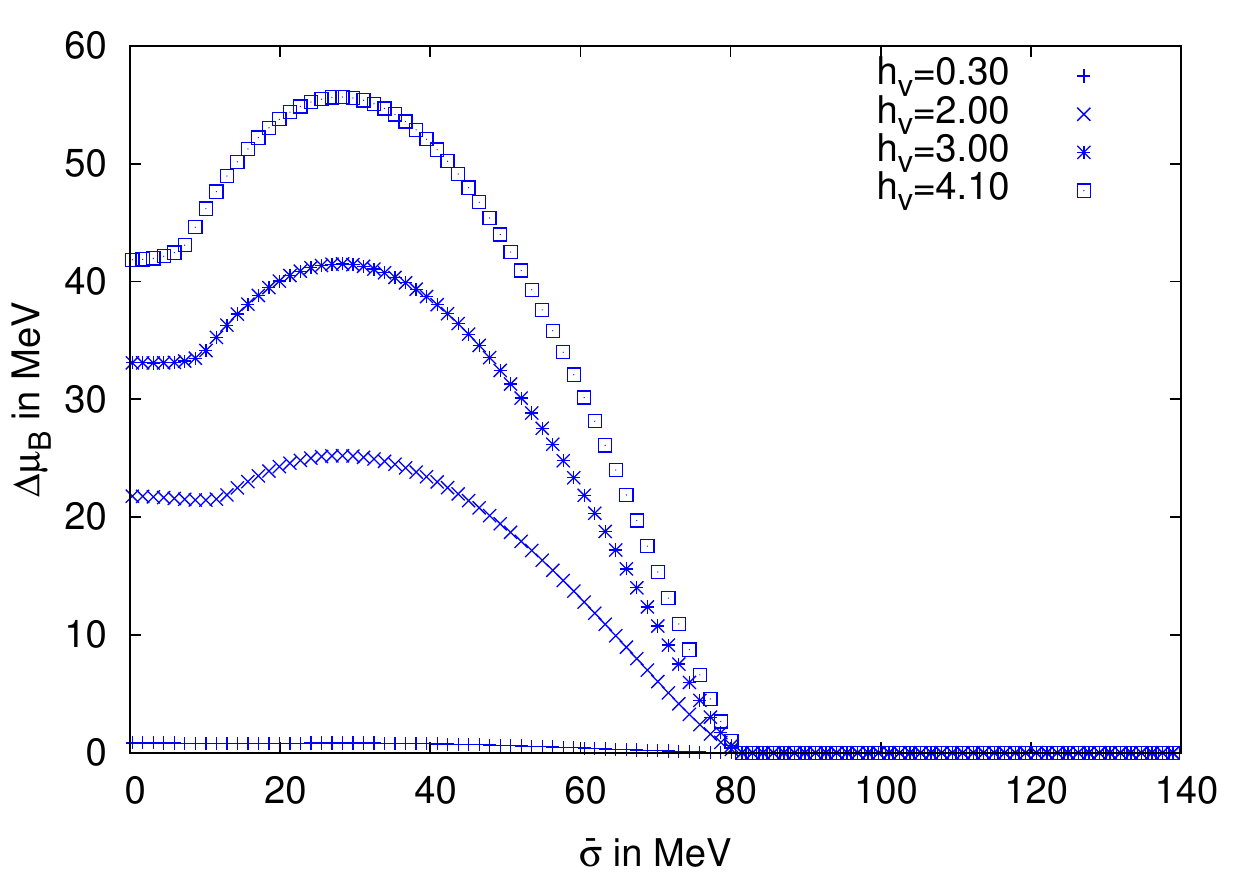}
\caption{(Color online) Solutions $\Delta\mu_B = \imag h_v \bar\omega_0$ of the $\omega$-gap equation at $m_0=820$\;MeV, $\mu_B=892$\;MeV
as a function of the chiral condensate for different values of the vector coupling $h_v$.}
\label{emfomega}
\end{figure}

The corresponding solutions for $\Delta\mu_B=\imag h_v\bar\omega_0$ 
 of the $\omega$-gap equation, cf. App.~\ref{Gap-equations}, as a function of $\bar\sigma$ are shown in
 Fig.~\ref{emfomega}.
These were all obtained at $\mu_B=892$\;MeV which corresponds to the nuclear-matter transition for the smallest vector coupling $h_v=0.30$.  
This illustrates how the first-order phase transition at $\mu_B^c$ gets
shifted to larger values of $\mu_B$, and weakened at the same time,
by an increasing vector coupling, due to a non-zero shift $\Delta\mu_B$ in the chemical potential. In order to achieve this desired effect it is crucial, however, to have a non-trivial solution with non-zero  $\Delta\mu_B=\imag h_v \bar\omega_0$  for $\bar\sigma$ values larger than the in-medium condensate at saturation density, $\bar\sigma(n_0) $, which is between 50 MeV and 69 MeV for the parameters used here (cf.~Fig.~\ref{emfcondensate1}). We also note that the bifurcation point in the $\omega $-gap equation stays put at around $\bar\sigma \simeq 80$ MeV for all values of the vector coupling which implies that $\mu_B^c$ can not be shifted any further, once $\bar\sigma(n_0) $ reaches this value, because $\Delta\mu_B(\bar\sigma(n_0)) $ remains at zero regardless of the size of the vector coupling from then on.

\begin{figure}[h]
\includegraphics[width=0.47 \textwidth]{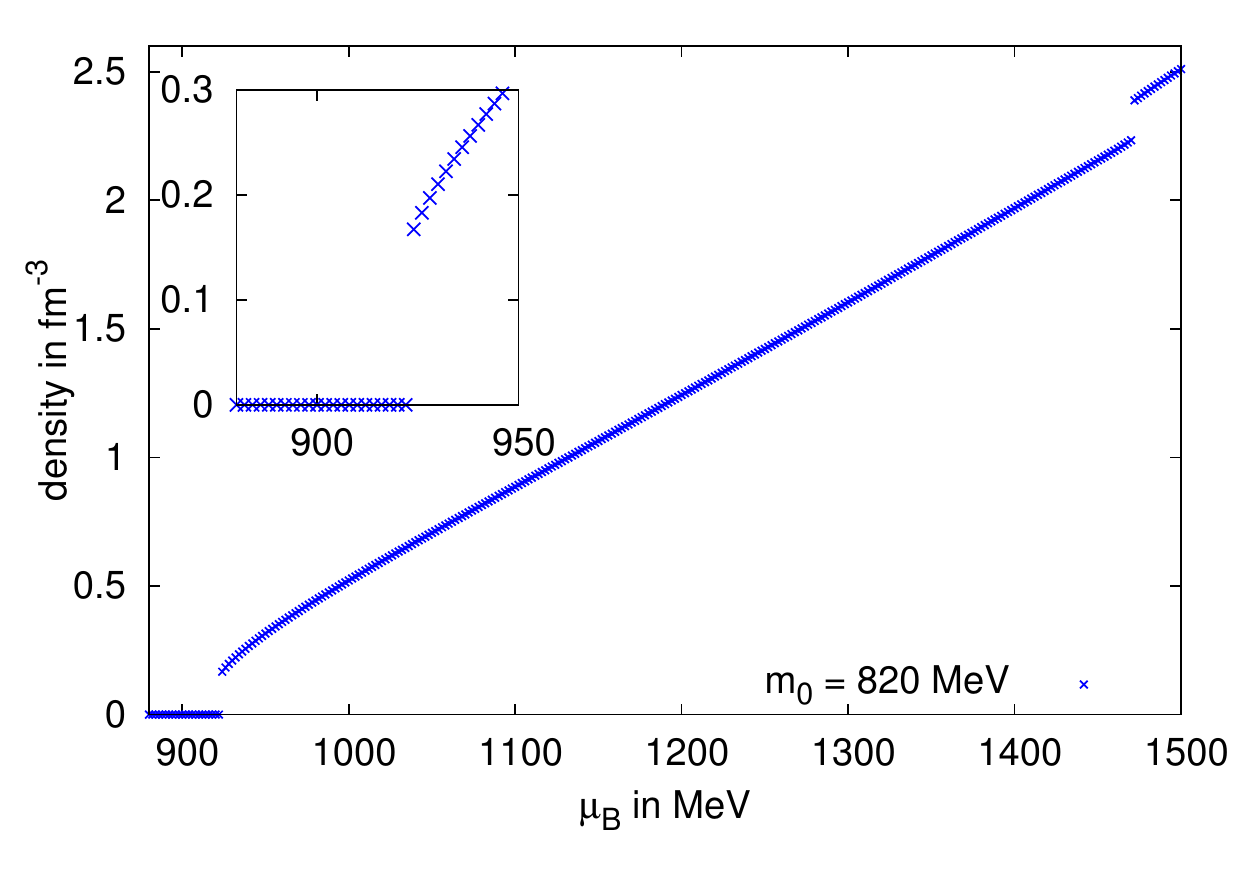}
\caption{(Color online) Baryon number density as a function of baryon chemical potential $\mu_B$ for $m_0=820$\;MeV and $h_v=4.10$.}\label{emfdens}
\end{figure}

Fig.~\ref{emfdens} shows the baryon density as a function of $\mu_B$ for $m_0=820$\;MeV and $h_v=4.10$, i.e. with the parameters for a phenomenologically correct  eMF description of the nuclear-matter transition. At $\mu_B^c=923$\;MeV the density jumps discontinuously from zero to $n_0\simeq0.16\;\text{fm}^{-3}$, the saturation density of symmetric nuclear matter. The second discontinuous transition (not shown in Fig.~\ref{emfcondensate1}) is observed at $\mu_B \simeq 1472 $ MeV. It occurs at a density of around 13.7 times the saturation density and would thus hardly be relevant for the equation of state of nuclear matter in neutron stars. Mesonic fluctuations might well change this, however, as they tend to bring both transitions much closer together as we will see in the next subsection.

\begin{figure}[h]
\centering
\includegraphics[width=0.47 \textwidth]{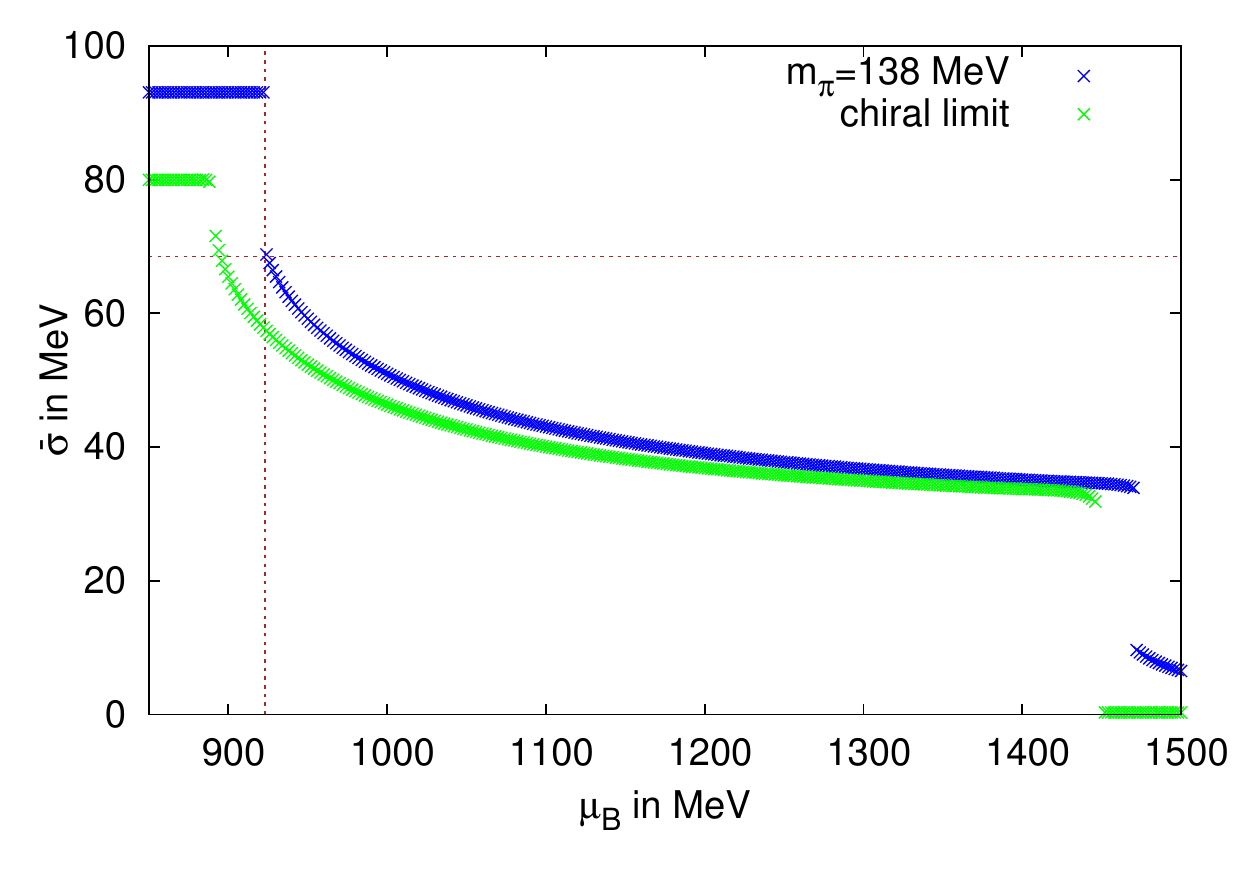}
\caption{(Color online) The chiral condensate as a function of $\mu_B$ for the nuclear-matter and for the chiral phase transition of the model, with physical pion mass and in the chiral limit.}\label{emfcondensate2}
\end{figure}

That the second transition within the high-density phase is indeed the chiral transition is demonstrated in  Fig.~\ref{emfcondensate2} which shows the chiral condensate over the baryon chemical potential again: 
The upper blue curve corresponds to the result of Fig.~\ref{emfcondensate1} with correct nuclear-matter parameters, the pion mass was thereby adjusted to its physical value, $m_\pi=138$\;MeV. The lower green curve was obtained with the same parameters except for the explicit chiral symmetry-breaking parameter which was set to $c=0$ for the chiral limit with $m_\pi =0$. The fact that  $\bar\sigma$ then drops to zero in the second transition allows to unambiguously identify this as the chiral first-order phase transition of the model inside the high baryon-density phase. These results also allow to deduce the parameters of the nuclear-matter transition in the chiral limit, with $\bar\sigma_0 \simeq 80 $ MeV, $E_b\simeq 47 $ MeV,   $ n_0 \simeq 0.064 $ fm$^{-3}$ and  $\bar\sigma(n_0) \simeq 0.89 \, \bar\sigma_0$.
\begin{figure}[t]
\centering
\includegraphics[width=0.47 \textwidth]{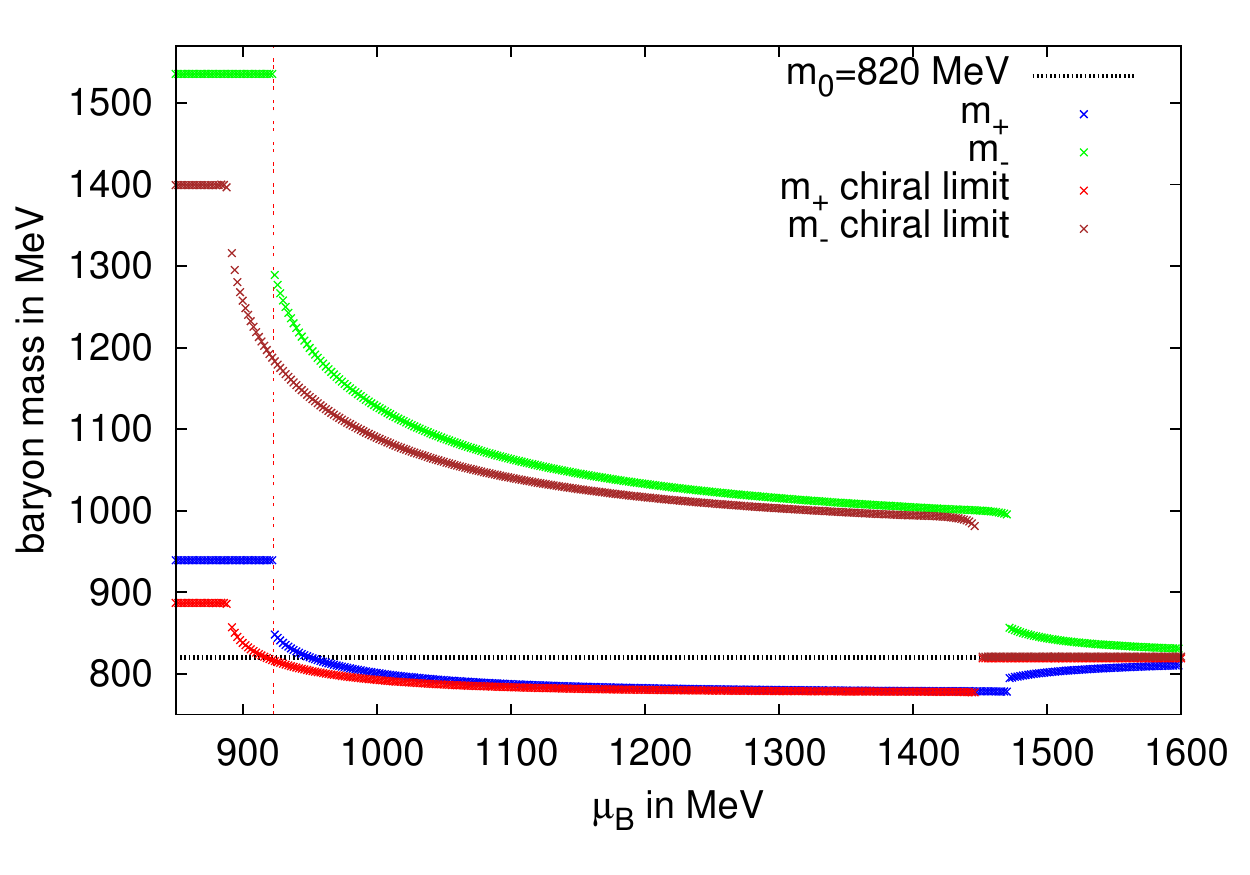}
\caption{(Color online) Masses of the nucleons and their parity partners at $m_0=820$\;MeV for physical pion mass and in the chiral limit.}\label{emfmasses}
\end{figure}
In Fig.~\ref{emfmasses} the corresponding masses of the nucleons  and their parity partners as a function of $\mu_B$ are shown for both, the physical pion mass
and in the chiral limit as well. It is evident that they become  degenerate
at large $\mu_B$, after the second phase transition.
For a physical pion mass the splitting of the baryon masses stays finite after the chiral phase transition and then smoothly tends to zero with  $m_\pm \to m_0$. This confirms the mechanism through which chiral
symmetry is realized in the parity doublet model. 
Rather than yielding vanishing masses in the chiral limit the baryon masses become degenerate at $m_\pm=m_0$, as observed in the lattice QCD simulations upon chiral symmetry restoration \cite{Glozman:2012iu,Glozman:2012fj,Glozman:2012hw}.
%%%%%%%%%%%%%%%%%%%%%%%%%%%%%%%%%%%%%%%%%%%%%%%%%%%%%%%%%%%%%%%%%%%%%%%%%%%%%%%%%%%%%%%%%%%%%%%%%%%%%%%%%%%%%%%%%%%%%%%%%%%%%%%%%%%%%%%%%%%%%%%%%%%%%%%%%%%%%%

\subsection{FRG Results with Mesonic Fluctuations}\label{fullRGsection}
%%%%%%%%%%%%%%%%%%%%%%%%%%%%%%%%%%%%%%%%%%%%%%%%%%%%%%%%%%%%%%%%%%%%%%%%%%%%%%%%%%%%%%%%%%%%%%%%%%%%%%%%%%%%%%%%%%%%%%%%%%%%%%%%%%%%%%%%%%%%%%%%%%%%%%%%%%%%%%
In this section we present our results obtained from the full FRG flow, the solutions of the full flow equation discretized on a grid in field space \cite{Schaefer:2006sr}, including the fluctuations from collective mesonic excitations as per Eq.~(\ref{eq:bosflow}).
\begin{figure}[h]
\centering
\includegraphics[width=0.47 \textwidth]{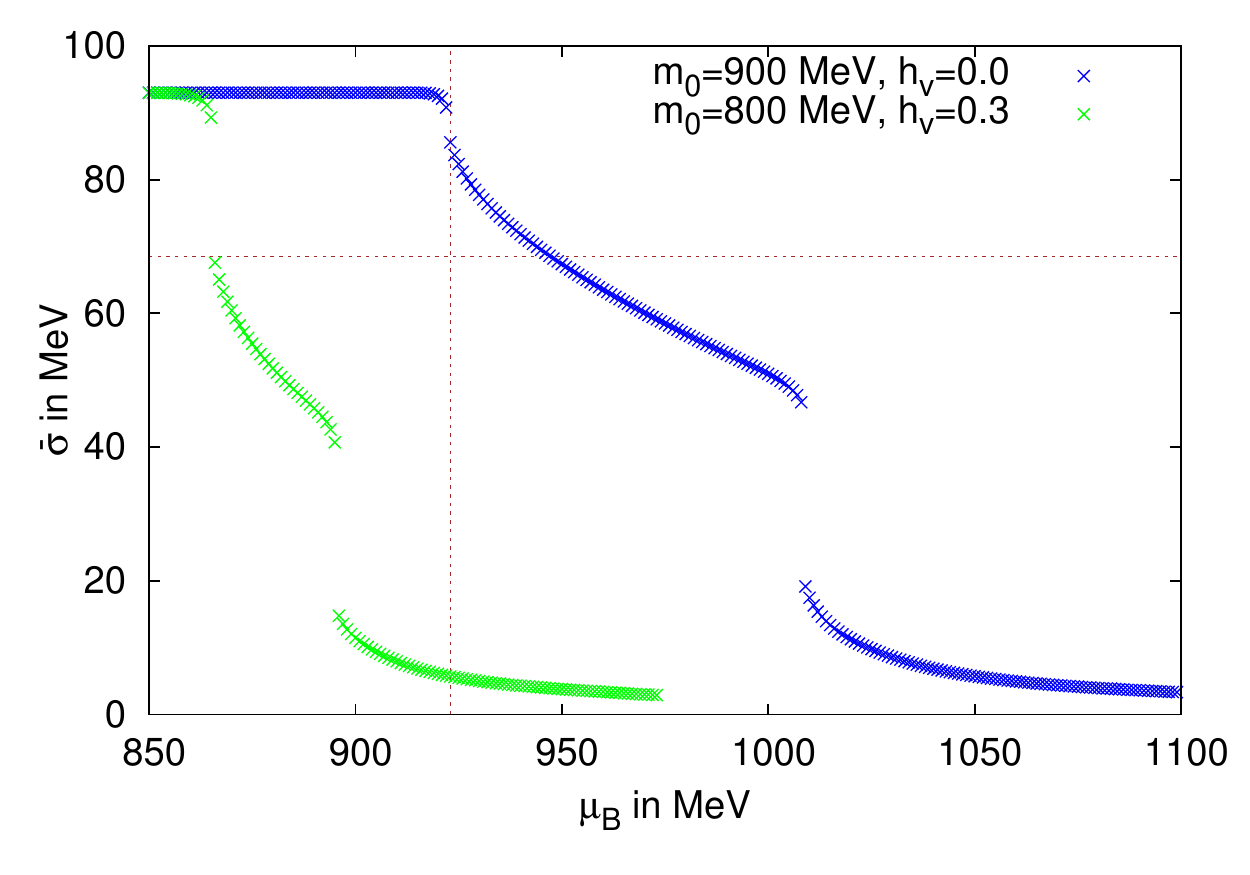}
\caption{(Color online) The chiral condensate as a function of $\mu_B$ for the nuclear-matter and the chiral phase transition with $m_0= 800$\;MeV and parameters fixed to the in-medium condensate (horizontal dotted line), and $m_0=900$\;MeV with parameters fixed to the binding energy per nucleon (vertical dotted line).}\label{fullRGcondensate}
\end{figure}

In Fig.~\ref{fullRGcondensate} the resulting expectation value of chiral condensate $\bar\sigma$ is shown as a function of baryon chemical potential $\mu_B$. The two representative data sets correspond to two different values for $m_0=800$ MeV and $900$ MeV with different vector couplings $h_v=0.3$ and $h_v = 0$, respectively. In both sets the other parameters were adjusted to reproduce the physical pion and baryon masses.
For $m_0=800$ MeV the nuclear-matter and chiral phase transitions occur at $\mu_B^{c\, (n)} \simeq 866$ MeV and
$\mu_B^{c\, (\chi)} \simeq 896$ MeV, respectively. For $m_0=900$~MeV these values are $\mu_B^{c\, (n)} \simeq 923$ MeV for nuclear matter and $\mu_B^{c\, (\chi)} \simeq 1009$~MeV for the chiral transition. The corresponding nuclear-matter saturation densities for the two parameter sets are  $n_0 \simeq 0.10\;$fm$^{-3}$ with $m_0=800$~MeV and
$n_0 \simeq 0.01\;$fm$^{-3}$ with $m_0=900$ MeV. They are both much smaller than the phenomenological $n_0 \simeq 0.16\;$fm$^{-3}$  which appears to be due to missing density contributions from the omega meson.
As can be seen in this figure we generally observe that one can either reproduce the physical binding energy per nucleon (as indicated by the dotted red vertical line) or the correct in medium condensate (the dotted red horizontal line)
but not both at the same time.
\begin{figure}[H]
\centering
\includegraphics[width=0.47 \textwidth]{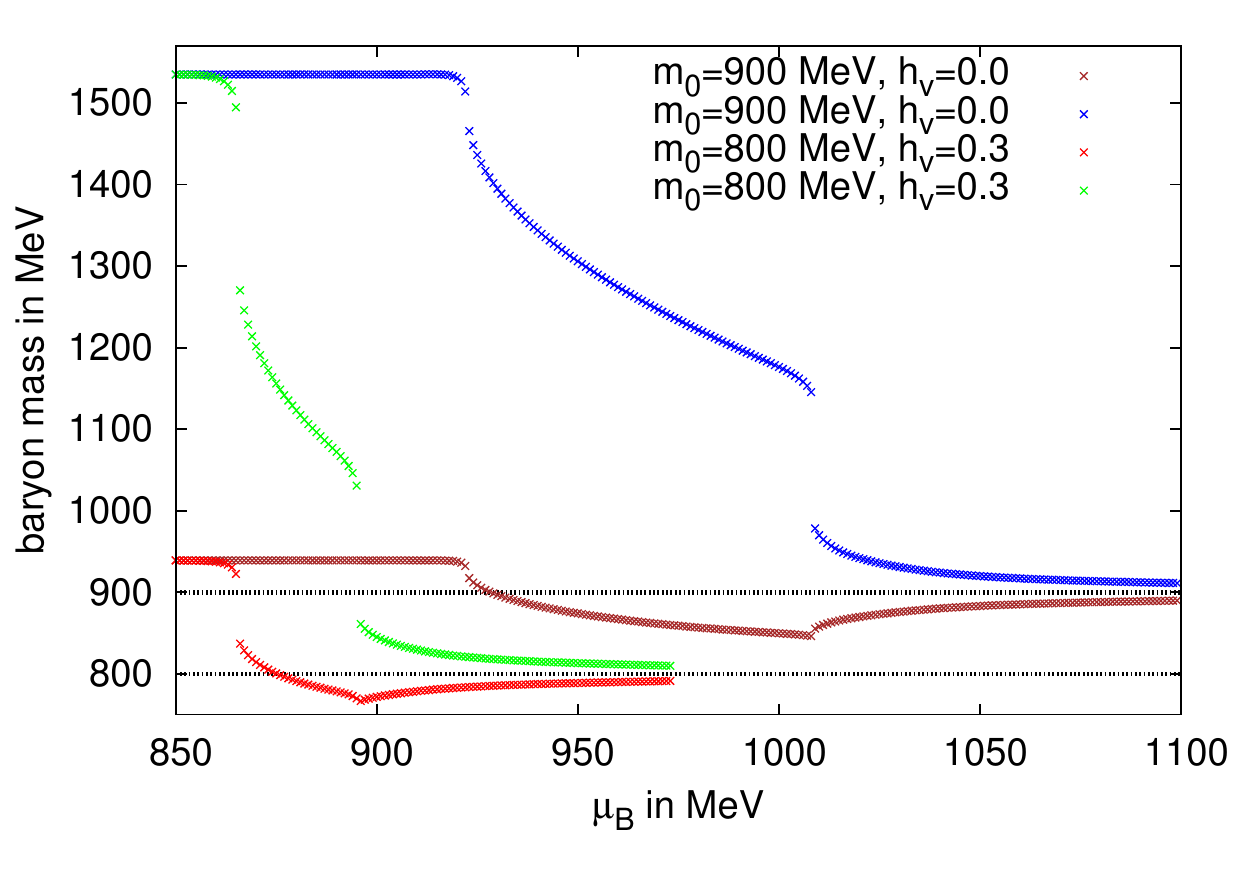}
\caption{(Color online) Masses of the nucleons and their parity partners at $m_0=800$\;MeV and $900$\;MeV with physical pion masses, full FRG results with parameters as in Fig.~\ref{fullRGcondensate}.}\label{fullRGmasses}
\end{figure}

\begin{figure*}
  \centering \subfloat[]{
\includegraphics[width=0.48\textwidth]{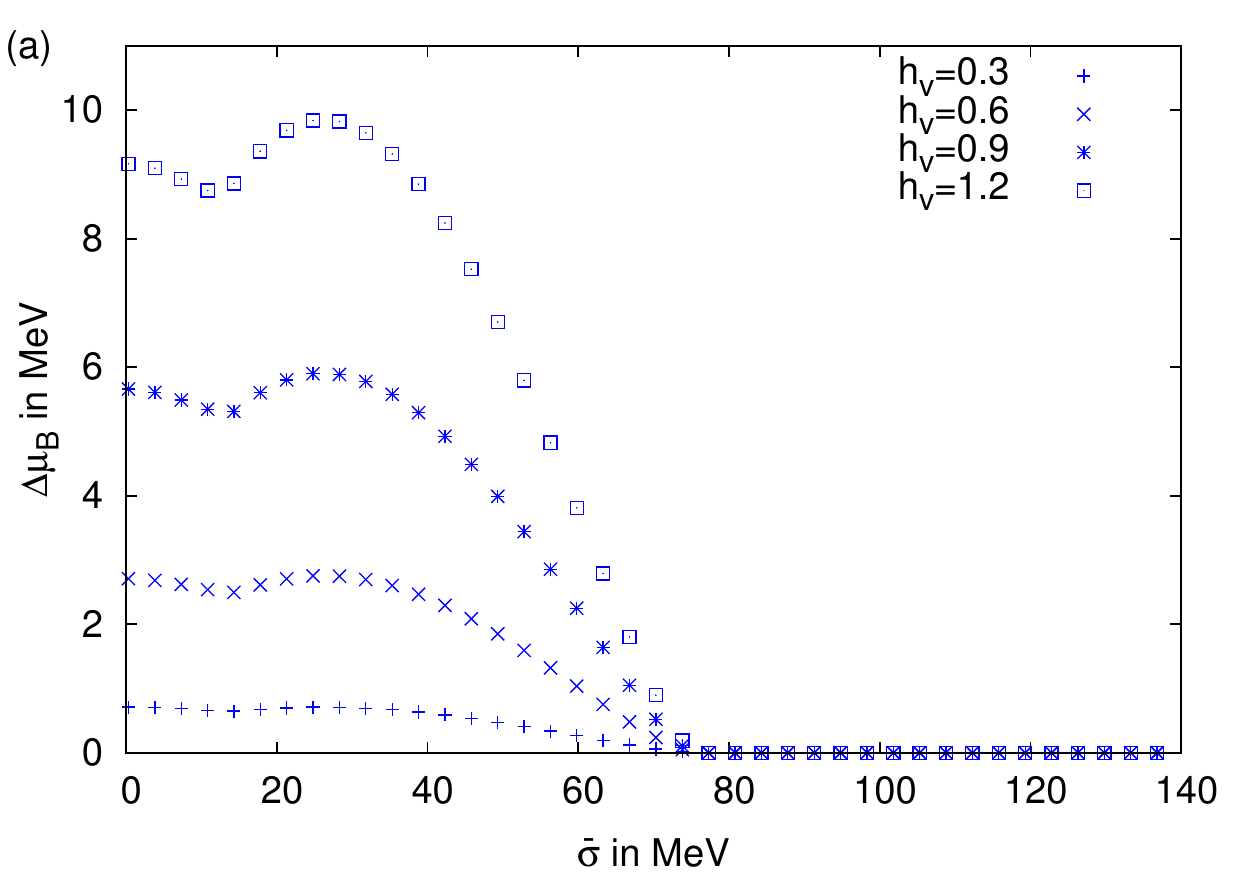}
    \label{fullRGomega}} \hfill \subfloat[]{\includegraphics[width=0.48
\textwidth]{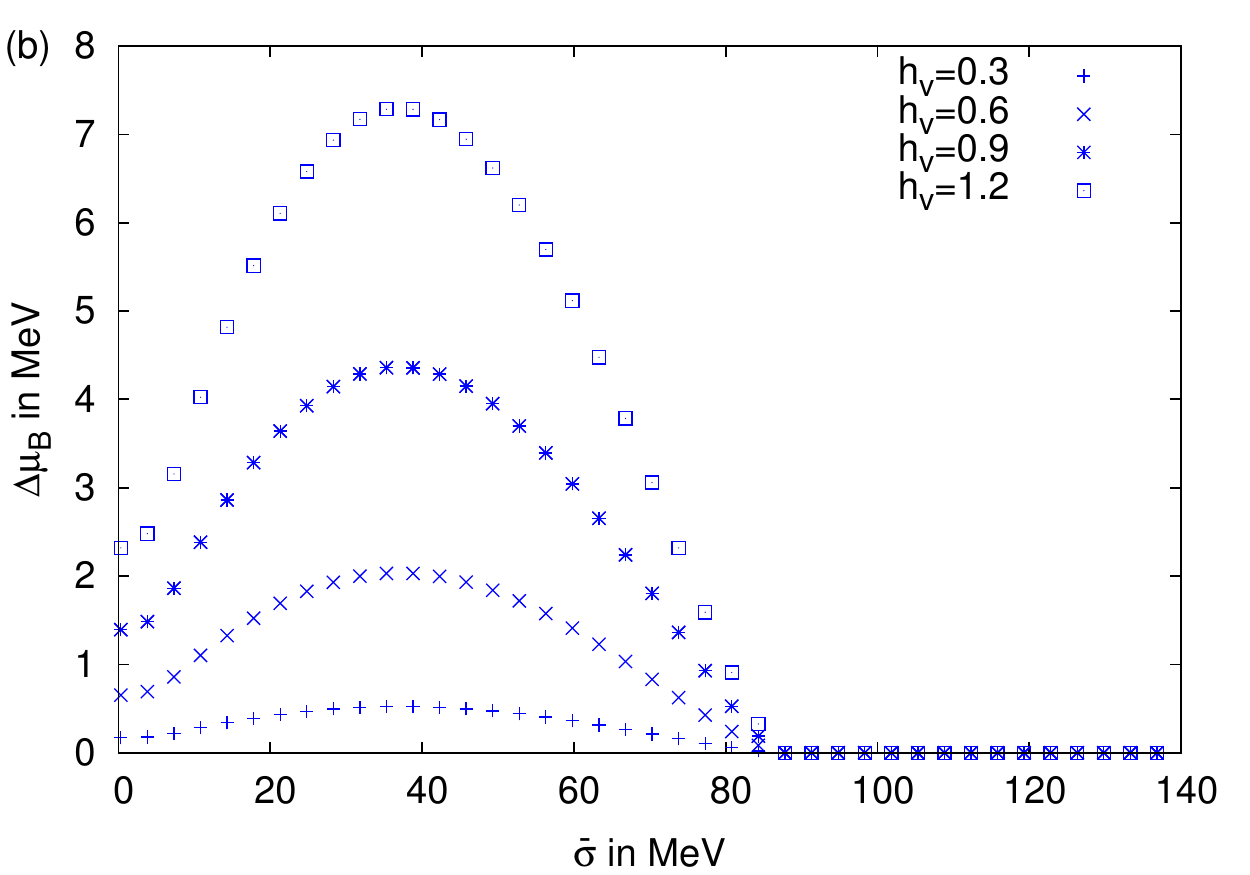}
\label{fullRGomega2}}
\caption{(Color online) Solutions of the $\omega$-gap equation for different vector couplings $h_v$ with $m_0= 800 $ MeV (left) and $m_0 = 900$ MeV (right), and values for the chemical potential at the respective zero-temperature nuclear-matter transitions $\mu_B=866$~MeV and $\mu_B=923$~MeV, as in Fig.~\ref{fullRGcondensate}. }\label{fullRGomegas}
\end{figure*}

The corresponding baryon masses are shown for the same two parameter sets in
Fig.~\ref{fullRGmasses}. As before, they both tend to $m_0$ from above and below beyond the chiral transition which is generally closer to the liquid-gas transition than in the eMF results as mentioned above.

%\begin{figure}[b]
%\includegraphics[width=0.4 \textwidth]{fullRG_omega.pdf}
%\caption{Solutions of the $\omega$-gap equation at $m_0=900$\;MeV, $\mu_B=923$\;MeV for different values of the vector coupling $h_v$.}\label{fullRGomega}
%\end{figure}

%\begin{figure}[t]
%\includegraphics[width=0.4 \textwidth]{fullRG_omega_2.pdf}
%\caption{Solutions of the $\omega$-gap equation at $m_0=800$\;MeV, $\mu_B=866$\;MeV for different values of the vector coupling $h_v$.}\label{fullRGomega2}
%\end{figure}

The reason why, in contradistinction to mean-field and eMF calculations, the repulsive vector interaction is ineffective in shifting the nuclear-matter transition with full mesonic fluctuations becomes clear from Figs.~\ref{fullRGomega} and \ref{fullRGomega2}. In these figures we plot the shifts $\Delta\mu_B$ in the baryon chemical potential from the solutions to the $\omega$-gap equations over the chiral condensate for various vector couplings with the two values of $m_0$ used in Fig.~\ref{fullRGcondensate}. 
%In both cases the bifurcation points in the $\omega$-gap equations are not larger than the in-medium %condensates $\bar\sigma(n_0)$ at the respective saturation densities.%
%The regimes of non-vanishing%
%$\Delta\mu_B= \imag h_v\bar\%omega_0$ do not extend beyond the position of the minimum in the $\sigma %$-%direction of the effective p%otential at saturation density, and the larger vector couplings thus %essentially have no effect o%n the nuclear-matter transition. %
%Non-vanishing $\bar\omega_0$ values hence only change the form of the local potential away from t%he %minimum, but not the grand potential which is evaluated at $\bar\sigma_0$ and $\bar\sigma(n_0)$ on either %side of the transition.

For $m_0=900$~MeV the bifurcation point in the $\omega$-gap equation always lies below the value of the in-medium condensate $\bar\sigma(n_0)$ at the respective saturation density. In this case the regime of non-vanishing
$\Delta\mu_B= \imag h_v\bar\omega_0$ does not extend beyond the position of the minimum in the $\sigma$-direction of the effective potential at saturation density, and larger vector couplings thus have no effect on the nuclear-matter transition. In the case with $m_0=800$~MeV, even
though the bifurcation point at $\bar\sigma \simeq 75$ MeV lies slightly above $\bar\sigma(n_0)\simeq 69$~MeV, the repulsive vector coupling
only has a very minor influence on the in-medium condensate. For example, without vector coupling we obtain with this parameter set $\bar\sigma(n_0)\simeq 67$~MeV which we can shift to the phenomenological $\bar\sigma(n_0)\simeq 69$~MeV by increasing the vector coupling from $h_v=0.0$ to $h_v=0.3$. Larger vector couplings will not lead to any substantial further shift in the nuclear-matter transition. In addition, we have verified that the bifurcation points in the $\omega$-gap equation remain where they are in Figs.~\ref{fullRGomega} and \ref{fullRGomega2} also for much larger values of the vector coupling $h_v$ than those shown there.
We conclude that the non-vanishing $\Delta\mu_B$ values essentially only change the form of the local potential away from the minimum. Depending on the chiral nucleon-mass parameter $m_0$ they do either not at all or not significantly influence the grand potential which is evaluated at $\bar\sigma_0$ and $\bar\sigma(n_0)$ on either side of the transition.

%%%%%%%%%%%%%%%%%%%%%%%%%%%%%%%%%%%%%%%%%%%%%%%%%%%%%%%%%%%%%%%%%%%%%%%%%%%%%%%%%%

\subsection{Finite Temperature Results}

Since the flow equation for the effective potential in the parity-doublet model
of the previous section is already formulated for finite temperatures it is in principle straightforward to obtain the phase diagram of the model in the $(T,\mu_B)$-plane. In this subsection we present first results from the full RG flow at finite temperature (and chemical potential). As the impact of the repulsive 
vector-meson interaction can be neglected in full RG calculations, as we have seen, it is not included here, i.e.~we use $h_v=0$ in the following. These results are only meant to give a rough qualitative estimate of the critical
temperatures  $T_c^{(n)}$ for the nuclear-matter and  $T_c^{(\chi)}$ for the chiral transition. 
\begin{figure*}
  \centering \subfloat[]{
\includegraphics[width=0.48\textwidth]{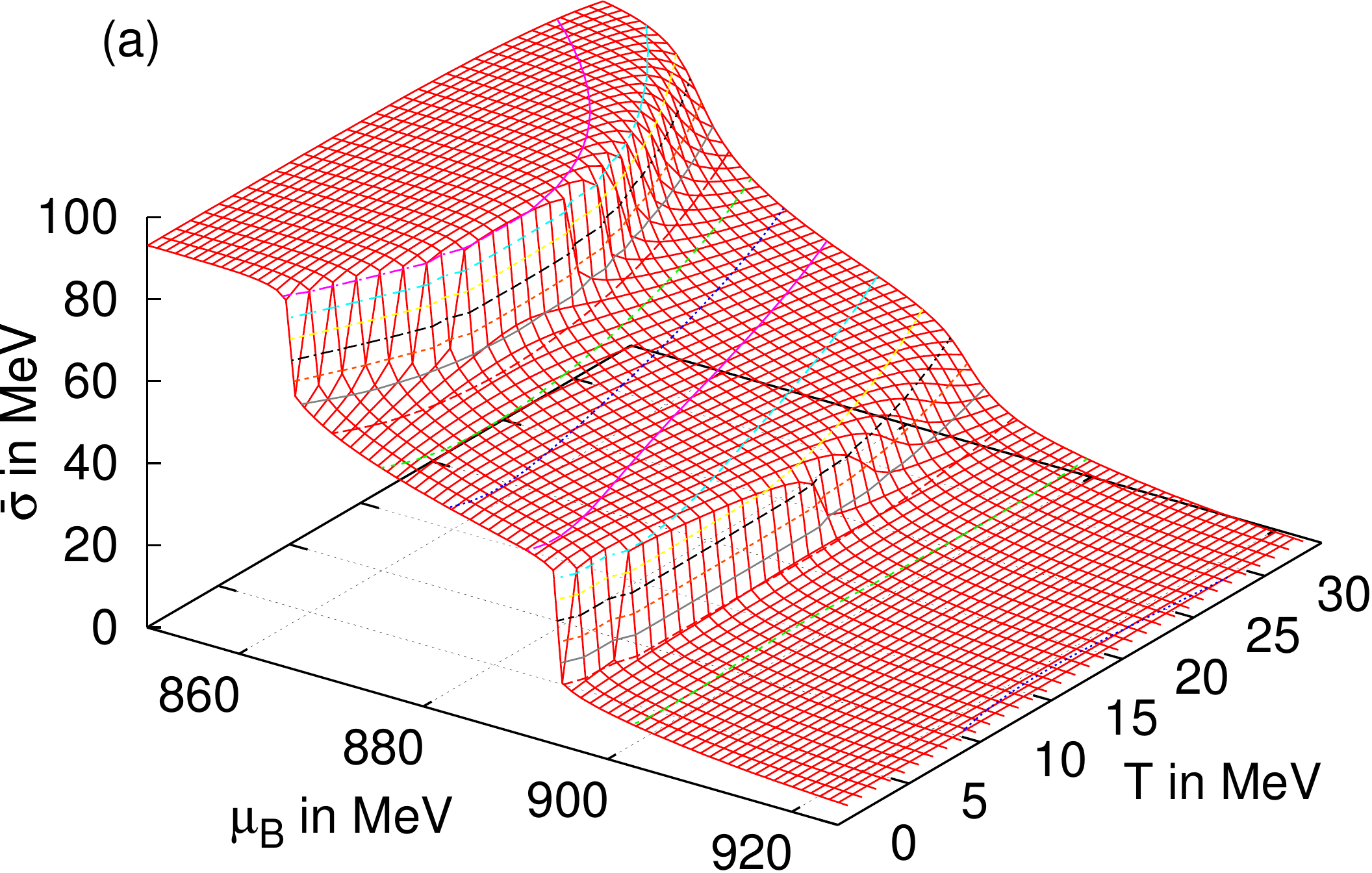}
    \label{fullRGfiniteTm0800}} \hfill \subfloat[]{\includegraphics[width=0.48
\textwidth]{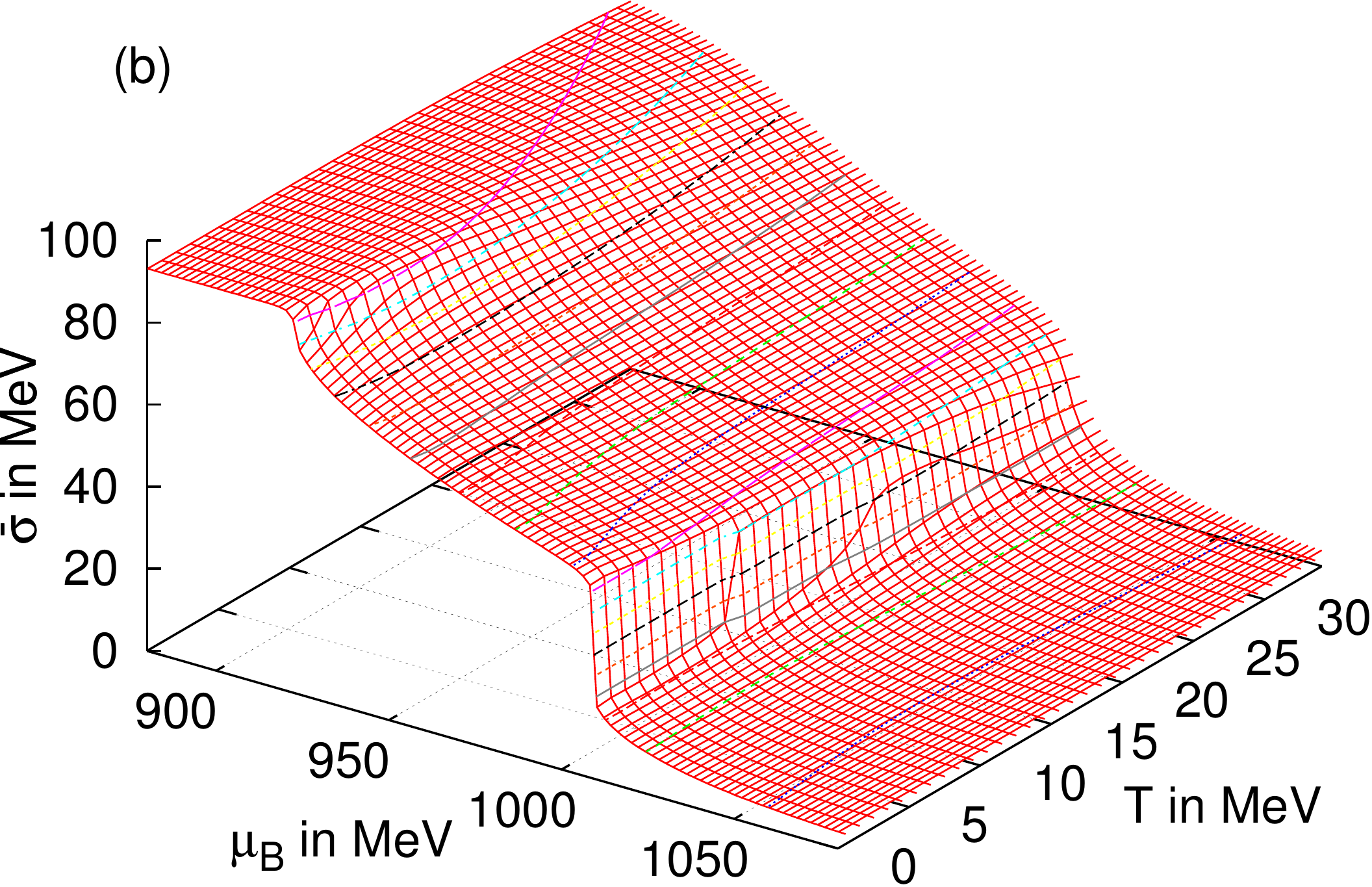}
\label{fullRGfiniteTm0900}}
  \caption{(Color online) Chiral condensate over chemical potential $\mu_B$  and temperature $T$ from the full FRG flow for the two different parameter sets with $m_0 = 800$ MeV (left) and $m_0=900$ MeV (right) from Sec.~\ref{fullRGsection}, both with $h_v=0$. }\label{fullRGfiniteTs}
 %   \hfill\textcolor{white}{.} 
\end{figure*}
We have chosen the two parameter sets from Sec.~\ref{fullRGsection} (cf. Fig.~\ref{fullRGcondensate}), both with $h_v=0$ here, to perform the finite temperature calculations presented in this subsection. The results are summarized in Figs.~\ref{fullRGfiniteTm0800} and \ref{fullRGfiniteTm0900} as well as in Figs.~\ref{phasediagram800} and \ref{phasediagram900}.
\begin{figure*}
  \centering \subfloat[]{
\includegraphics[width=0.48
\textwidth]{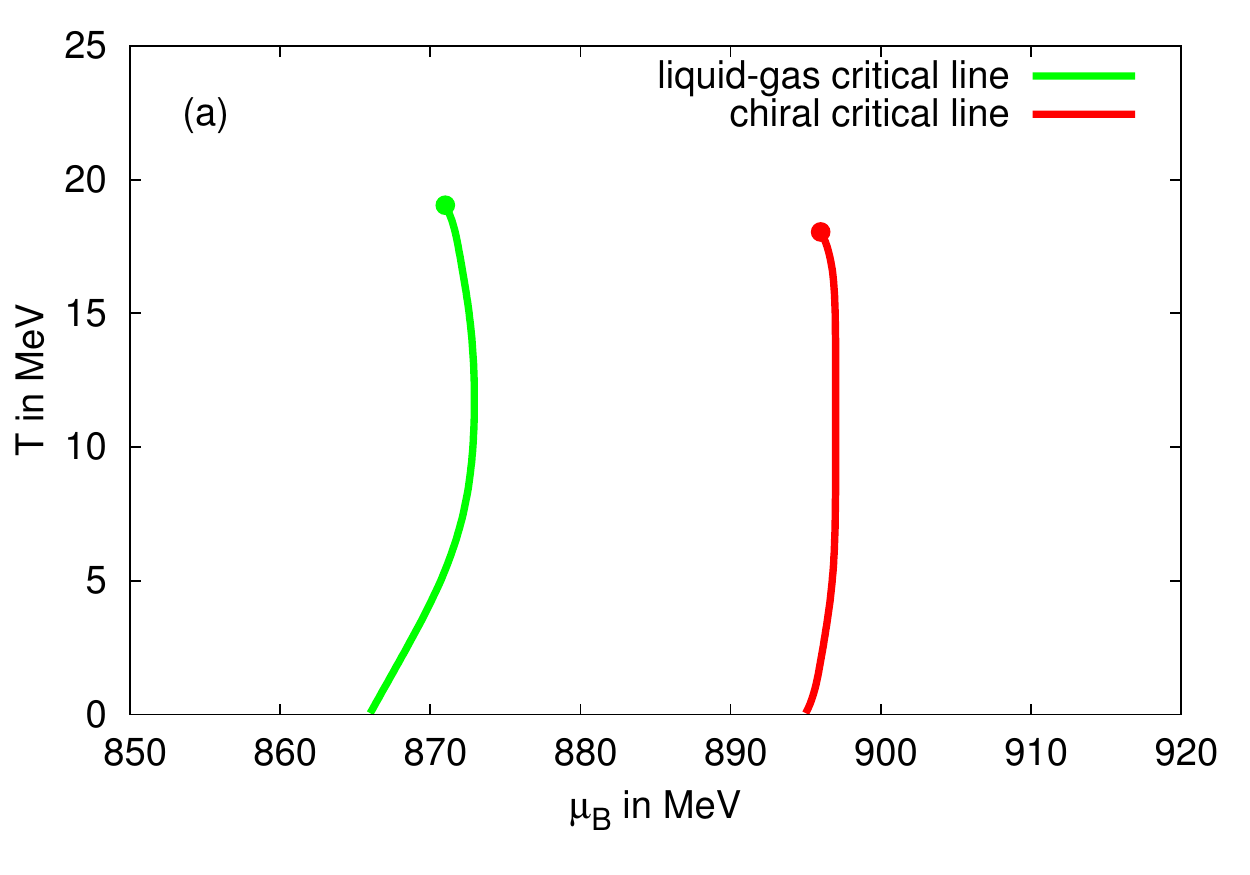}
\label{phasediagram800}} \hfill \subfloat[]{\includegraphics[width=0.48\textwidth]{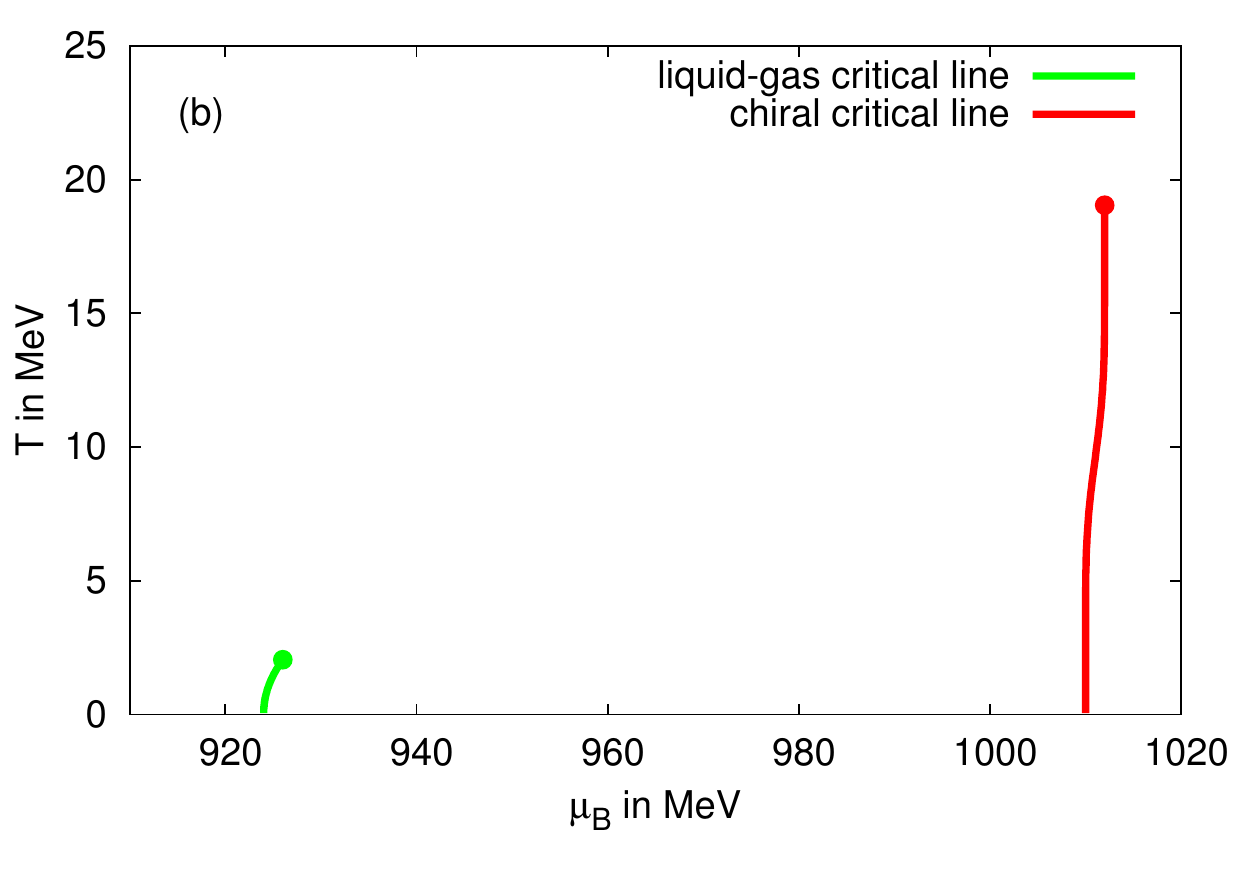}
\label{phasediagram900}}
  \caption{(Color online) First-order lines with the corresponding critical endpoints in the $(T,\mu_B)$-plane for the liquid-gas (green) and chiral (red) phase transitions using the same data sets as in Figs.~\ref{fullRGfiniteTm0800} and \ref{fullRGfiniteTm0900} for $m_0 = 800$ MeV (left) and $m_0=900$ MeV (right). }\label{fullRGphasediagrams} 
 %   \hfill\textcolor{white}{.} 
\end{figure*}

%\begin{figure}[b]
%  \includegraphics[width=0.5\textwidth]{finiteT_nm_m0=800.pdf}
%\caption{Chiral condensate over chemical potential $\mu_B$  and temperature $T$ for the results from the full RG flow with $m_0=800$\;MeV.}\label{fullRGfiniteTm0800}
%\end{figure}

In Fig.~\ref{fullRGfiniteTm0800} the chiral condensate for $m_0=800$\;MeV (with $h_v =0 $  and the other parameters as in Sec.~\ref{fullRGsection}) is plotted over the  $(T,\mu_B)$-plane. As expected, both first-order phase transitions, the nuclear matter and the chiral transition, get weaker with increasing temperature and eventually become continuous. The corresponding critical lines in the $(T,\mu_B)$-plane are shown in Fig.~\ref{phasediagram800}.
The location of the critical endpoint (CEP) of the nuclear-matter transition can be estimated to be at  $T_c^{(n)}\sim 19$\;MeV and $\mu_B^{c\,(n)} \simeq 871$\;MeV. In comparison, that of the chiral CEP is around  $T_c^{(\chi)}\sim 18$\;MeV and $\mu_B^{c\,(\chi)} \simeq 896$\;MeV. 

%\begin{figure}[b]
%\includegraphics[width=0.5\textwidth]{finiteT_nm_m0=900.pdf}
%\caption{Chiral condensate over chemical potential $\mu_B$  and temperature $T$ for the results from the full RG flow with  $m_0=900$\;MeV.}\label{fullRGfiniteTm0900}
%\end{figure}

The analogous plots of the chiral condensate and the critical lines for the parameter set with
$m_0=900$\;MeV are shown in Figs.~\ref{fullRGfiniteTm0900} and \ref{phasediagram900}.
In this case the nuclear-matter transition is much weaker in the first place. It thus also turns continuous already at a very low temperature of 
$T_c^{(n)} \simeq 2$\;MeV with  $\mu_B^{c\,(n)} \simeq 925 $\;MeV.
We estimate the location of the chiral CEP to now be around  $T_c^{(\chi)}\sim 19$\;MeV and $\mu_B^{c\,(\chi)} \simeq 1012$\;MeV.

Relative to their corresponding zero temperature transitions at $\mu_B^{c\,(n)} \simeq 866$\;MeV and  $\mu_B^{c\,(\chi)} \simeq 895$\;MeV for $m_0 = 800$~MeV, or  $\mu_B^{c\,(n)} \simeq 923$\;MeV and  $\mu_B^{c\,(\chi)} \simeq 1015$\;MeV for $m_0 = 900$~MeV, especially the nuclear-matter CEPs tend to appear at somewhat larger $\mu_B$ values. The chiral first-order transitions from $T=0$ to $T_c^{(\chi)}$ basically almost follow lines of constant $\mu_B$, especially for $m_0 = 900$\;MeV. 

The slope of the first-order lines in the $(T,\mu_B)$-plane is determined by a
Clausius-Clapeyron equation \cite{Kogut:2004su},
\begin{equation}
  \frac{dT_c}{d\mu_c} = - \frac{\Delta n}{\Delta s} \,.
\end{equation}
For the discontinuity in the number density we have $\Delta n > 0 $ with increasing $\mu_B$ across the first-order lines in both cases. For an order-disorder transition one would expect the entropy per particle to increase, and with $\Delta n>0$ hence also the discontinuity in the entropy density to be larger than zero, i.e. $\Delta s > 0$. This is the typical behavior of the chiral transition line in mean-field studies. Here, $\Delta s$ in the chiral transition with mesonic fluctuations and explicit symmetry breaking appears to be very small. It may well be positive in the chiral limit.

More interestingly, because the entropy per particle decreases from the gaseous to the liquid phase, it is possible to have $\Delta s < 0$ in the liquid-gas transition of nuclear matter despite the fact that the number density increases across the transition. This is what we obtain for the nuclear-matter transition in the parity-doublet model with mesonic fluctuations for $m_0 = 900$ MeV and at low temperatures also for $m_0 =800$ MeV. It is a genuine effect of mesonic fluctuations that they tend to change the sign in the discontinuity of the entropy density as compared to mean-field studies. It is known from the quark-meson model where they change the mean-field chiral transition into a transition to bound quark matter \cite{Schaefer:2004en}. The same effect turns the relativistic analogue of a Chandrasekhar-Clogston transition inside the pion condensation phase at finite isospin chemical potential, as observed at mean-field with $\Delta s > 0$, into a first-order transition to a stable Sarma phase with $\Delta s <0$, when mesonic fluctuations are included \cite{Kamikado:2012bt,Boettcher:2014xna}. This would be analogous to a partially polarized phase in ultracold Fermi gases at unitarity. For the nuclear-matter transition in QCD it appears to be rather unusual to have $\Delta s < 0$. It is not observed in chiral effective field theory for example \cite{Fiorilla:2011sr}. Since $\Delta n$ tends to be too small in our FRG results with mesonic fluctuations, however, especially for $m_0 = 900$ MeV, the slope of the nuclear-matter transition line might still turn out to be negative with fluctuations as well, if a more realistic $\Delta n$, with $\Delta n \simeq 0.16 $ fm$^{-3}$ at $T=0$, is sufficient for the entropy density to also increase across the transition (leading to $\Delta s > 0$) despite the fact that the entropy per particle should decrease.

%%%%%%%%%%%%%%%%%%%%%%%%%%%%%%%%%%%%%%%%%%%%%%%%%%%%%%%%%%%%%%%%%%%%%%%%%%%%%%%%%%%%%%%%%%%%%%%%%%%%%%%%%%%%%%%%%%%%%%
\section{Conclusions}
\label{sec:conclusion}
We conclude, that the inclusion of a heavy parity partner in a chiral baryon-meson model such as the parity-doublet model within an FRG framework allows for
a simultaneous description of the liquid-gas transition of nuclear matter together with a chiral first order transition inside the high baryon-density phase. The quantitative properties of symmetric nuclear matter are well reproduced in the extended mean-field approximation without collective mesonic fluctuations. Including mesonic fluctuations does not change the qualitative conclusion of the existence of two distinct first-order phase transitions which  can be identified as a liquid-gas transition of nuclear matter and the chiral phase transition at which the nucleons become degenerate with their parity partners.

First calculations at finite temperature (and chemical potential) provide the general features of the phase diagram of the model with the two first-order lines ending in two distinct critical endpoints. The inclusion of mesonic fluctuations thereby has effects, especially on the liquid-gas transition, that are known from quark-meson models. As compared to (extended) mean-field studies they lead to a sign-change in the slope of the critical line indicating a sign-change in the discontinuity of the entropy density. 

In contrast to the case with purely baryonic fluctuations, however, the repulsive iso-scalar vector meson interaction turns out to be inefficient in adjusting the binding energy per nucleon of symmetric nuclear matter. As a result, the FRG treatment of the parity-doublet model with mesonic fluctuations does not provide a quantitatively fully successful phenomenological description of the nuclear-matter transition at this point. A possible remedy which has proven to work for the chiral Walecka model \cite{Floerchinger:2012xd,Drews:2013hha} would be to parametrize the effective potential in the vacuum and to consider only thermal fluctuations. In a longer term one might embed the parity-doublet model into a (Polyakov-)quark-meson-baryon model to include the effects of fluctuating light quark degrees of freedom in the chirally restored and deconfined phase and at the initial microscopic scales.

\subsection*{Acknowledgements}
This work is supported by the Helmholtz International Center for FAIR  within the LOEWE program of the State of
Hesse. N.S. is supported by grant ERC-AdG-290623.
\appendix
\section{Gap Equations}
    \label{Gap-equations}
%%%%%%%%%%%%%%%%%%%%%%%%%%%%%%%%%%%%%%%%%%%%%%%%%%%%%%%%%%%%%%%%%%%%%%%%%%%%%%%%%%%%%%%%%%%%%%%%%%%%%%%%%%%%%%%%
%%%%%%%%%%%%%%%%%%%%%%%%%%%%%%%%%%%%%%%%%%%%%%%%%%%%%%%%%%%%%%%%%%%%%%%%%%%%%%%%%%%%%%%%%%%%%%%%%%%%%%%%%%%%%%%%
    For the grand potential we need to minimize the resulting effective action with respect to the $\sigma$-meson field. The $\omega$-meson is not included as a fluctuating field here, but treated in a stationary phase approximation with a complex saddle point for the $\omega_0$-integration as mentioned in the main text. For both, the minimum of the effective potential in the $\sigma$-direction 
    and the (purely imaginary) saddle point in $\omega_0$, we thus require the partial derivatives of $U_0$, where 
    $U_0 = U_k$ for $k\to 0$,  with respect to $\sigma$ and  $\omega_0$ to vanish when $\sigma=\bar\sigma$ and $\omega_0=\bar\omega_0$.
     The corresponding two gap equations are of the form,
			\begin{equation}\label{sigmagap}
				\frac{\partial}{\partial \sigma}\left(\int \text{d}k \, \partial_kU_k\right)\bigg{|}_{\sigma=\bar\sigma,\omega_0=\bar\omega_0}= c\,,
			\end{equation}
and
	                \begin{equation}\label{omegagap}
				\frac{\partial}{\partial \omega_0}\left(\int \text{d}k\,  \partial_kU_k\right)\bigg{|}_{\sigma=\bar\sigma,\omega_0=\bar\omega_0}=-m_\omega^2\bar\omega_0\, .
			\end{equation}
Vice versa Eqs.~(\ref{omegagap}) and (\ref{sigmagap}) define the expectation values of the meson fields $\bar\sigma$ and $\bar\omega_0$.

The gap equation for $\omega_0$ can be derived from the 
fermionic RG flow in Eq.~(\ref{eq:fermionmirrorflow}). The derivative with respect to $\omega_0$ is thereby  equivalent to a derivative with respect to the baryon chemical potential $\mu_B$,
			\begin{equation}
				\frac{\partial}{\partial\omega_0}=-\imag h_v\frac{\partial}{\partial\mu_B}
			\end{equation}
For $T\to0$ the gap equation for $\omega_0$ can be obtained analytically in closed form from the expression,
			\begin{equation}\label{omegagapcold}
				\frac{\partial}{\partial \omega_0}\int \text{d}k\, \partial_kU_k\bigg{|}_{\bar\sigma,\bar\omega_0}
				=\imag h_v\frac{4}{6\pi^2}\sum_{\pm}\frac{k_{F}^{\pm4}\gamma^{\pm}(k_F^{\pm})}{|\partial_k\epsilon_k^{\pm}(k)|_{k=k_F^{\pm}}|}\, ,
			\end{equation}
with 
			\begin{equation}
				\gamma^{\pm}(k)=\frac{2(k^2+m_0-\epsilon_k^{\pm 2})+\sigma^2(h_1^2+h_2^2)}{\epsilon_k^{\pm}(\epsilon_k^{\mp 2}-\epsilon_k^{\pm 2})} \, .
			\end{equation}
Here, $k^\pm_F$ are the Fermi momenta defined by
			\begin{equation}
				\epsilon_k^\pm(k_F^{\pm})=\tilde \mu_B.
			\end{equation}
                        Eq.~(\ref{omegagapcold}) also allows us to obtain the baryon number density,
                        \begin{equation}
                          n_B = - \frac{\partial U_0}{\partial\mu_B} = \frac{1}{\imag h_v} \, \frac{\partial}{\partial \omega_0} \, \int \text{d}k\,  \partial_kU_k\bigg{|}_{\bar\sigma,\bar\omega_0}  .
                        \end{equation}
With $n_b^\pm$ for the contributions to the number densities of nucleons and their parity partners separately we thus find,
			\begin{equation}
			n_B =\sum_{\pm}n_B^{\pm}= \frac{4}{6\pi^2}\sum_{\pm}\frac{k_{F}^{\pm4}\gamma^{\pm}(k_F^{\pm})}{|\partial_k\epsilon_k^{\pm}(k)|_{k=k_F^{\pm}}|}\, .
			\end{equation}

\bibliographystyle{bibstyle}
\bibliography{mirrormodel}
\end{document}